\documentclass[aps,superscriptaddress,prd,
onecolumn,
floatfix, 
nofootinbib,
amsmath,amssymb,amsfonts,longbibliography]{revtex4-2}

\usepackage[pdftex]{graphicx}
\usepackage[colorlinks,linkcolor=blue,anchorcolor=violet,citecolor=red]{hyperref}

\usepackage{float,color}
\usepackage{rsfso}
\usepackage{physics}
\usepackage{comment}
\newcommand{\anno}[1]{\textcolor{red}{#1}}

\newcommand{\al}{\alpha}
\newcommand{\pa}{\partial} 
\newcommand{\del}{\delta}
\newcommand{\bs}[1]{\boldsymbol{#1}}

\date{July 30, 2023} 

\begin{document}
\title{Entanglement partners and monogamy in de Sitter universes} 
%
\author{Yasusada Nambu}
\email{nambu@gravity.phys.nagoya-u.ac.jp}
\affiliation{Department of Physics, Graduate School of Science, Nagoya
  University, Nagoya 464-8602, Japan}
\author{Koji Yamaguchi}
\email{koji.yamaguchi@uwaterloo.ca}
\affiliation{Department of Applied Mathematics, University of
  Waterloo, Waterloo, Ontario, N2L 3G1,Canada}

\begin{abstract} 
  We investigate entanglement of local spatial modes defined by a
  quantum field in a de Sitter universe.  The introduced modes show
  disentanglement behavior when the separation between two regions
  where local modes are assigned becomes larger than the cosmological
  horizon. To understand the emergence of separability between these
  local modes, we apply the monogamy inequality proposed by
  S. Camalet. We embed the focusing bipartite mode defined by the
  quantum field in a pure four-mode Gaussian state, and identify its
  partner modes. Then applying a Gaussian version of the monogamy
  relation, we show that the external entanglement between the
  bipartite mode and its partner modes constrains the entanglement of
  the bipartite mode. Thus the emergence of separability of local
  modes in the de Sitter universe can be understood from the perspective     
  of entanglement monogamy.
\end{abstract}   

\maketitle
\section{Introduction} 
Cosmic inflation explains the origin of structures in our Universe by
preparing seeds of primordial fluctuations as quantum origin, vacuum
fluctuations of a quantum scalar field called inflaton. The
contribution of this quantum field to energy density functions
as a cosmological constant, leading to the accelerated
expansion of the Universe. During the rapid expansion of the
Universe, vacuum fluctuations receive parametric amplification, and
the resulting fluctuations evolve to become ``classical" seed
fluctuations causing gravitational instability and later
  forming the large scale structures
\cite{Kiefer2009}. Although this is an accepted scenario of
structure formation based on cosmic inflation in standard
  cosmology, the mechanism of ``quantum to classical transition" of
primordial fluctuation has not been well understood yet.

Entanglement is a key concept to differentiate quantum systems
  from classical ones and a crucial tool to investigate the quantum
  nature of the initial stage of our universe. In our previous
studies
\cite{Nambu2008,Nambu2009,Nambu2011,Nambu2013,Matsumura2018,Nambu2023a},
local oscillator modes defined from the quantum scalar field in a de
Sitter universe were investigated, and it was found that the initial
entangled state becomes separable; that is, two local modes A and B, which are
assigned at two spatial regions, become separable when their
separation exceeds the Hubble horizon scale.  This disentanglement
behavior can be explained as follows: the ``thermal noise" with the
Gibbons-Hawking temperature associated with the de Sitter horizon
breaks quantum correlations between two spatial regions, and
therefore, the entangled bipartite state of modes A and B becomes
separable. After these two modes become separable, only classical
correlations survive between them.

The mechanism of disentanglement can also be studied from the
property of multipartite entanglement.  The bipartite system AB is
defined as a subsystem of the total system, i.e., the field in the
entire universe.  Although the total system is assumed to be in a pure
state, modes AB are in a mixed state since they are correlated with
its complement $\overline{\text{AB}}$. It is always possible to find
 a subsystem in $\overline{\text{AB}}$ that purifies AB, which is
called the partner mode of AB.
Then, we can understand the disentanglement of AB as a result of
entanglement sharing between these modes and their partners. More concretely, the disentanglement of AB can be analyzed from the perspective
of entanglement monogamy in multipartite quantum systems
\cite{Coffman2000,Osborne2006,Hiroshima2007,Adesso2007a,Adesso2014,Dhar2016}. Monogamy
of entanglement is an intrinsic property of quantum correlations that
is not amenable to classical explanations. For the bipartite state AB
and its complement $\text{C}:=\overline{\text{AB}}$, the conventional
monogamy relation is expressed as the following inequality
 \begin{equation}
   E(\text{A:B})+E(\text{A:C})\le E(\text{A:BC}),
\label{eq:mono-conv}
 \end{equation}
 where $E(\text{X:Y})$ denotes a suitably chosen entanglement measure
 for a bipartite system XY. The inequality  restricts the
 amount of the bipartite entanglement $E(\text{A:B})$ as sharing of
 correlations in the tripartite system ABC. However, this inequality
 does not provide such a tight constraint as to derive a condition on
 the separability $E(\text{A:B})=0$ for multipartite Gaussian states
 \cite{Matsumura2018}. See also Appendix \ref{sec:appendix} where we
 review the monogamy property \eqref{eq:mono-conv} for a pure
 four-mode Gaussian state.

 A slightly different form of the monogamy relation was proposed by
 Camalet
 \cite{Camalet2017,Camalet2017a,Camalet2018,Zhu2020,Zhu2021}, which
 relates ``internal'' and ``external'' quantum correlations in
 multipartite states. Here, for a bipartite system AB of interest,
 the correlation between A and B is internal, while the one between AB
 and another subsystem X in the complementary system
 $\overline{\text{AB}}$ is external.  Based on assumptions of general
 correlation measures, a new kind of monogamy inequality was derived,
 which states that internal entanglement and external entanglement
 obey a trade-off relation. As a consequence, explicit forms of the
 monogamy inequality are obtained in terms of entanglement measures
 for finite-dimensional systems, such as qubits.
     
 In this paper, we investigate the entanglement behavior of local
 bipartite modes AB of a quantum field in the de Sitter universe from
 the viewpoint of the monogamy of entanglement. For this purpose, we
 identify partner modes that purify the bipartite modes AB by using
 the formalism proposed in
 \cite{Trevison2018b,Tomitsuka2019,Yamaguchi2020b,Yamaguchi2020a,Yamaguchi2020}. We
 then prove a Camalet-type trade-off relation between internal and
 external correlations for these modes, i.e., a monogamy relation on
 the entanglement between A and B, and the entanglement between AB and
 their partners. Based on these formalisms, we find that the emergence
 of separability between local modes in the de Sitter universe can be
 understood from the viewpoint of entanglement monogamy.

 The paper is organized as follows. In Sec. \ref{sec:2}, we introduce
 a quantum scalar field in the de Sitter universe and show the
 disentanglement behavior of local modes assigned at two spatial
 points. In Sec. \ref{sec:3}, we review our method of
   construction of the partner modes of the two local modes based on
 the formulas in
 \cite{Trevison2018b,Tomitsuka2019,Yamaguchi2020b,Yamaguchi2020a,Yamaguchi2020}. 
 In Sec. \ref{sec:4}, based on the result obtained in
 Sec. \ref{sec:3}, we examine Camalet's monogamy relation for Gaussian
 modes with which the emergence of separability in the de Sitter
 universe is
 analyzed. 
 Section \ref{sec:5} is devoted to summary and conclusion.  We adopt
 units of $c=\hbar=1$ throughout this paper.

\section{Scalar field and local modes}\label{sec:2}
To comprehend the behavior of the entanglement of quantum fields in
the de Sitter universe, we consider a minimally coupled massless
scalar field $\hat\phi$ in a (3+1)-dimensional flat
Friedmann-Lema\^{i}tre-Robertson-Walker (FLRW) universe. The scalar
field obeys the Klein-Gordon equation $\square\hat\phi=0$. The metric
of the FLRW universe with the conformal time $\eta$ and the comoving
coordinate $\bs{x}=(x,y,z)$ is
\begin{equation}
  ds^2=a_\text{sc}^2(\eta)(-d\eta^2+d\bs{x}^2), \label{eq:metric}
\end{equation}
where $a_\text{sc}(\eta)$ is the scale factor of the universe. We
will later fix its functional form as that which corresponds to the de
Sitter universe. The rescaled scalar field
$\hat\varphi=a_\text{sc}\hat\phi$ obeys the following field equation
\begin{equation}
\hat\varphi''-\left(\frac{a''_\text{sc}}{a_\text{sc}}+\del^{ij}\pa_i\pa_j\right)\hat\varphi=0,\quad i,j=x,y,z,
\end{equation}
where $'=d/d\eta$. We adopt this mode equation in a (3+1)-dimensional spacetime, but we assume that excitation propagates only in one spatial direction to simplify the analysis. 
Then the field operators of the massless scalar field are expressed as
\begin{align}
  &\hat\varphi(x)=\int_{-\infty}^\infty\frac{dk}{\sqrt{2\pi}}\,\hat\varphi_{k}\,e^{ikx},
    \quad
    \hat\varphi_{k}=f_k(\eta)\hat a_{k}+f_k^*(\eta)\hat a_{-k}^{\dag},\\
  &\hat\pi(x)=\int_{-\infty}^\infty\frac{dk}{\sqrt{2\pi}}\,\hat\pi_{k}
    \,e^{ikx},\quad
    \hat\pi_{k}=(-i)(g_k(\eta)\hat a_{k}-g_k(\eta)^*\hat a_{-k}^\dag),\\
  &[\hat a_{k_1},\hat a^\dag_{k_2}]=\del(k_1-k_2),\quad
    f_k''+\left(k^2-\frac{a_\text{sc}''}{a_\text{sc}}\right)f_k=0,\quad
    g_k=i\left(f_k'-\frac{a_\text{sc}'}{a_\text{sc}}f_k\right),\quad f_kg_k^*+f_k^*g_k=1,
\end{align}
where $\hat a_k$ and $\hat a_k^\dag$ are annihilation and creation
operators, and $\hat\pi(x)$ is the conjugate momentum of the field
variable $\hat\varphi(x)$. In terms of the Fourier components of the field
operators, the creation and the annihilation operators can be
represented as
\begin{equation}
  \hat
  a_{k}=g_k^*\hat\varphi_{k}+if_k^*\hat\pi_{k},\quad
  \hat a_{-k}^\dag =g_k\hat\varphi_{k}-if_k\hat\pi_{k}.
\end{equation}
We assume that the scalar field is in the vacuum state $\ket{\psi}$ associated with the annihilation operator
$\hat a_k$ such that
\begin{equation}
  \hat a_k\ket{\psi}=0.
\end{equation}
 The equal-time commutation relations for the field operators are given by
\begin{equation}
  [\hat\varphi(\eta,x),\hat\pi(\eta,y)]=i\del(x-y),\quad[\hat\varphi(\eta,x),\hat\varphi(\eta,y)]=[\hat\pi(\eta,x),\hat\pi(\eta,y)]=0.\label{eq:equal_time_ccr}
\end{equation}
Covariances of the field operators are calculated as
\begin{align}
 M_{11}(x,y)&:=\expval{\{\hat\varphi(x),\hat\varphi(y)\}}=\frac{2}{\pi}\int_0^{\infty}
         dk      |f_k|^2\cos(k(x-y)),\\
 M_{22}(x,y)&:=\expval{\{\hat\pi(x),\hat\pi(y)\}}=
   \frac{2}{\pi}\int_0^{\infty} dk |g_k|^2\cos(k(x-y)),\\
 M_{12}(x,y)&:=\expval{\{\hat\varphi(x),\hat\pi(y)\}}
    =\frac{1}{\pi}
    \int_{0}^{\infty}
    dk\,i(f_kg_k^*-f_k^*g_k)\cos(k(x-y)),
\end{align}
where $\expval{\,}$ denotes the expectation value with respect to the state $\ket{\psi}$. 
\subsection{Local modes}
We consider measurement of the field operators $\hat\varphi,
\hat\pi$ at spatial points $x_\text{A}$ and $x_\text{B}$. The measurement process can be represented as
the interaction between the field operators and dynamical variables of
the measurement apparatus such as Unruh-DeWitt detectors \cite{Birrell1984}. In the present analysis, we do not specify
details of the apparatus but just assume the interaction Hamiltonian
between the field operators and the apparatus has the following form:
\begin{equation}
  H_\text{int}=\sum_{j=\text{A,B}}\lambda_j(t)g_j(\hat q_D,\hat
  p_D)\otimes\int dx (w_{1j}(x)\hat\varphi(x)+w_{2j}(x)\hat\pi(x)),
\end{equation}
where $g_j$ is a function of canonical variables of the measurement
apparatus $(\hat q_D,\hat p_D)$, $w_{1j}(x), w_{2j}(x)$ are window functions defining a spatial
local mode of the field at $x_\text{A,B}$, and $\lambda_j(t)$ is a
switching function of the interaction. In the present analysis, we do
not treat details of measurement protocols but only pay attention
to the behavior of the local modes of the quantum field introduced by
the window functions.

Let us introduce local operators at $x_j$ ($j=\text{A,B}$) using a
window function $w_j(x)=w(x-x_j)$ as
\begin{align}
 &\hat q_j:=\int_{-\infty}^\infty dx
   w_j(x)\hat\varphi(x)=
   \int_{-\infty}^{\infty} dk\, \hat\varphi_{k}\,
   e^{ikx_j}w_k, \label{eq:local-mode1}\\
   &\hat p_j:=\int_{-\infty}^\infty dx
     w_j(x)\hat\pi(x)=\int_{-\infty}^{\infty} dk\, \hat\pi_{k}\,
   e^{ikx_j}w_k, \label{eq:local-mode2}
\end{align}
where the window function is assumed to be localized around $x_j$\anno{,} and $w_k$ denotes the Fourier component of the window function:
\begin{equation}
  w_k=\frac{1}{\sqrt{2\pi}}\int_{-\infty}^{\infty}dx\, w(x)e^{ikx},\quad w_k=w_{-k}^*.
\end{equation}
We require that the window function is fixed so that the local operators $(\hat{q}_j,\hat{p}_j)$ define independent modes. In other words, they satisfy the canonical commutation relations given by
\begin{align}
  &[\hat q_i,\hat p_j]=i\int_{-\infty}^\infty dk\,
    e^{ik(x_i-x_j)}|w_k|^2\equiv i\del_{ij}, \label{eq:comm}\\
  &[\hat q_i,\hat q_j]=[\hat p_i,\hat p_j]=0.
\end{align}
Note that these commutators are independent of the state of the quantum field.
Covariances for the local operators are
\begin{align}
  c_1(i,j)&:=\expval{\{\hat q_i,\hat q_j\}}
            =4\int_{0}^{\infty} dk |w_k|^2|f_k|^2\cos k\Delta_{ij}, \\
  c_2(i,j)&:=\expval{\{\hat p_i,\hat p_j\}}
    =4\int_{0}^{\infty} dk |w_k|^2|g_k|^2\cos k\Delta_{ij}, \\
  c_3(i,j)&:=\expval{\{\hat q_i,\hat p_j\}}
    =2 \int_0^\infty
            dk|w_k|^2\,i(f_kg_k^*-f_k^*g_k)\cos(k\Delta_{ij}),
\end{align}
where $\Delta_{ij}:=x_i-x_j$.

\begin{figure}[H]
  \centering
   \includegraphics[width=0.8\linewidth]{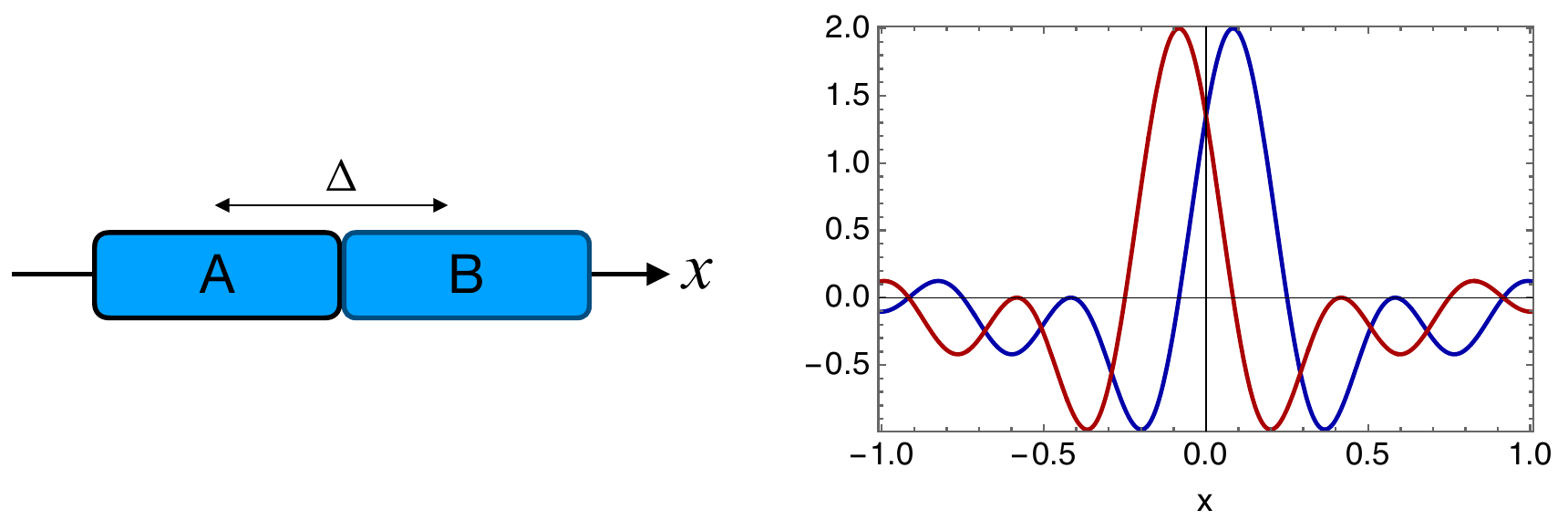}
   \caption{Left panel: setup of spatial regions A and B. Right panel: the 
     window functions for A: $w(x+\Delta/2)$ and B: $w(x-\Delta/2)$
     with $k_0/k_c=0.2$. Although the two window functions overlap,
     local modes A and B are independent and well-defined as the commutation
     relation \eqref{eq:comm} is satisfied.}
  \label{fig:setup}
\end{figure}

\paragraph{Window function}
We adopt a $k$-top hat window function in this study:
$w_k=w_0\,\theta(k_c-|k|)\theta(|k|-k_0),~k_c\ge k_0$, where $k_c$ is
the infrared (IR) cutoff corresponding to the total system size
(comoving size of the total universe) and $k_c$ is the ultraviolet
(UV) cutoff defining the size of localized modes. This type of a
window function was adopted in the stochastic approach to inflation
\cite{Starobinsky1986}, which is a phenomenological treatment of long
wavelength quantum fluctuations in the de Sitter universe, and this
method describes dynamics of the quantum inflaton field as a classical
stochastic variable obeying a Langevin equation.  The normalization
$w_0$ is determined by \eqref{eq:comm}:
\begin{equation}
  \del_{ij}=2w_0^2\int_{k_0}^{k_c}dk
  \cos(k\Delta_{ij})=2w_0^2\frac{\sin(k_c\Delta_{ij})-\sin(k_0\Delta_{ij})}{\Delta_{ij}}.
  \label{eq:norm}
\end{equation}
For $\Delta_\text{AA}=\Delta_\text{BB}=0$, Eq. \eqref{eq:norm}
provides the normalization of the window function that is determined as
$w_0^2=1/(2(k_c-k_0))$.
For $\Delta_\text{AB}\neq0$, Eq. \eqref{eq:norm} provides $
\sin(k_c\Delta_\text{AB})-\sin(k_0\Delta_\text{AB})=0$ which determines
\begin{equation}
  \Delta_\text{AB}=\frac{(2n-1)\pi}{k_c+k_0}, \frac{(2n)\pi}{k_c-k_0},\quad
  n=1,2,\dots. 
\end{equation}
As a value of $\Delta_\text{AB}$, we adopt the following in our analysis:
\begin{equation}
  |\Delta_\text{AB}|=\frac{\pi}{k_c+k_0}=:\Delta.
\end{equation}
The quantity $\Delta$ represents the distance between adjacent two local regions A and B
with $x_\text{B}-x_\text{A}=\Delta$ (Fig. \ref{fig:setup}). The $\Delta$
also represents the size of each local region.
The spatial profile of the window function is given by
\begin{equation}
  w(x)=\frac{2}{\sqrt{2\pi}}\int_{k_0}^{k_c}dk\,
  w_0e^{-ikx}=\frac{1}{\sqrt{\pi(k_c-k_0})}\frac{\sin(k_c x)-\sin(k_0x)}{x}. 
\end{equation}
\noindent
Covariances for the local operators are calculated as
\begin{align}
  &c_1(\Delta)=\frac{2}{k_c-k_0}\int_{k_0}^{k_c} dk |f_k|^2\cos k\Delta
    =\frac{2}{1-\del}\int_\del^1dz |f_k|^2\cos\left(\frac{\pi z}{1+\del}\right), \label{eq:c1}\\
  &c_2(\Delta)=\frac{2}{k_c-k_0}\int_{k_0}^{k_c} dk |g_k|^2\cos k\Delta
    =\frac{2}{1-\del}\int_\del^1dz |g_k|^2\cos\left(\frac{\pi z}{1+\del}\right), \label{eq:c2}\\
  &c_3(\Delta)=\frac{1}{k_c-k_0}\int_{k_0}^{k_c}
    dk\, i(f_kg_k^*-f_k^*g_k)\cos(k\Delta)
   =\frac{1}{1-\del}\int_\del^1dz\,
    i(f_kg_k^*-f_k^*g_k)\cos\left(\frac{\pi z}{1+\del}\right), \label{eq:c3}
\end{align}
where $z=k/k_c, \del=k_0/k_c$. The parameter $\del$ represents the size
of the local region normalized by the total system size:
$\del=(k_0\Delta/\pi)(1-k_0\Delta/\pi)^{-1}$. 

The covariance matrix of the bipartite system AB defined by
$(\hat q_\text{A},\hat p_\text{A},\hat q_\text{B},\hat p_\text{B})$ is given by
\begin{equation}
  \bs{m}_\text{AB}=
  \begin{bmatrix}
    a_1&a_3&c_1&c_3\\
    a_3&a_2&c_3&c_2\\
    c_1&c_3&a_1&a_3\\
    c_3&c_2&a_3&a_2
  \end{bmatrix},
  \label{eq:mAB}
\end{equation}
where $a_i:=c_i(\Delta=0)$ for $i=1,2,3$. Owing to the homogeneity of
the universe represented by the metric \eqref{eq:metric}, the
covariance matrices of each mode A and B have the same components;
i.e., the bipartite system AB is in a symmetric Gaussian
state. Symplectic eigenvalues of the covariance matrix
$\bs{m}_\text{AB}$ are calculated as
\begin{align}
  &(\nu_1)^2=a_1a_2-a_3^2+c_1c_2-c_3^2+|a_1c_2+a_2c_1-2a_3c_3|,\\
  &(\nu_2)^2=a_1a_2-a_3^2+c_1c_2-c_3^2-|a_1c_2+a_2c_1-2a_3c_3|.
\end{align}
The state represented by the covariance matrix \eqref{eq:mAB} is
physical, i.e., positive-semidefinite, if and only if
$1\le\nu_2\le\nu_1$.

The partially transposed covariance matrix, which is obtained by
reversing the sign of mode B's momentum, has the following two
symplectic eigenvalues:
\begin{align}
  &(\tilde\nu_1)^2=a_1a_2-a_3^2-c_1c_2+c_3^2+|(a_1c_2-a_2c_1)^2+4(a_1c_3-a_3c_3)(a_2c_3-a_3c_2)|^{1/2},
  \\
  &(\tilde\nu_2)^2=a_1a_2-a_3^2-c_1c_2+c_3^2-|(a_1c_2-a_2c_1)^2+4(a_1c_3-a_3c_3)(a_2c_3-a_3c_2)|^{1/2}.
\end{align}
Based on the positivity criterion of the partially transposed
covariance matrix for a bipartite Gaussian state
\cite{Peres1996,Horodecki1997,Simon2000}, the negativity gives a
measure of entanglement between modes A and B, which is defined as
\cite{Vidal2002a,Plenio2005}
\begin{equation}
  N_\text{A:B}:=\frac{1}{2}\mathrm{max}\left(\frac{1}{\tilde\nu_2}-1,0\right).
\end{equation}
The modes A and B are entangled if $N_\text{A:B}>0$, while the modes A
and B are separable if $N_\text{A:B}=0$. 
\subsection{Entanglement of local modes in the de Sitter universe}
\begin{figure}[H] 
  \centering
  \includegraphics[width=0.9\linewidth]{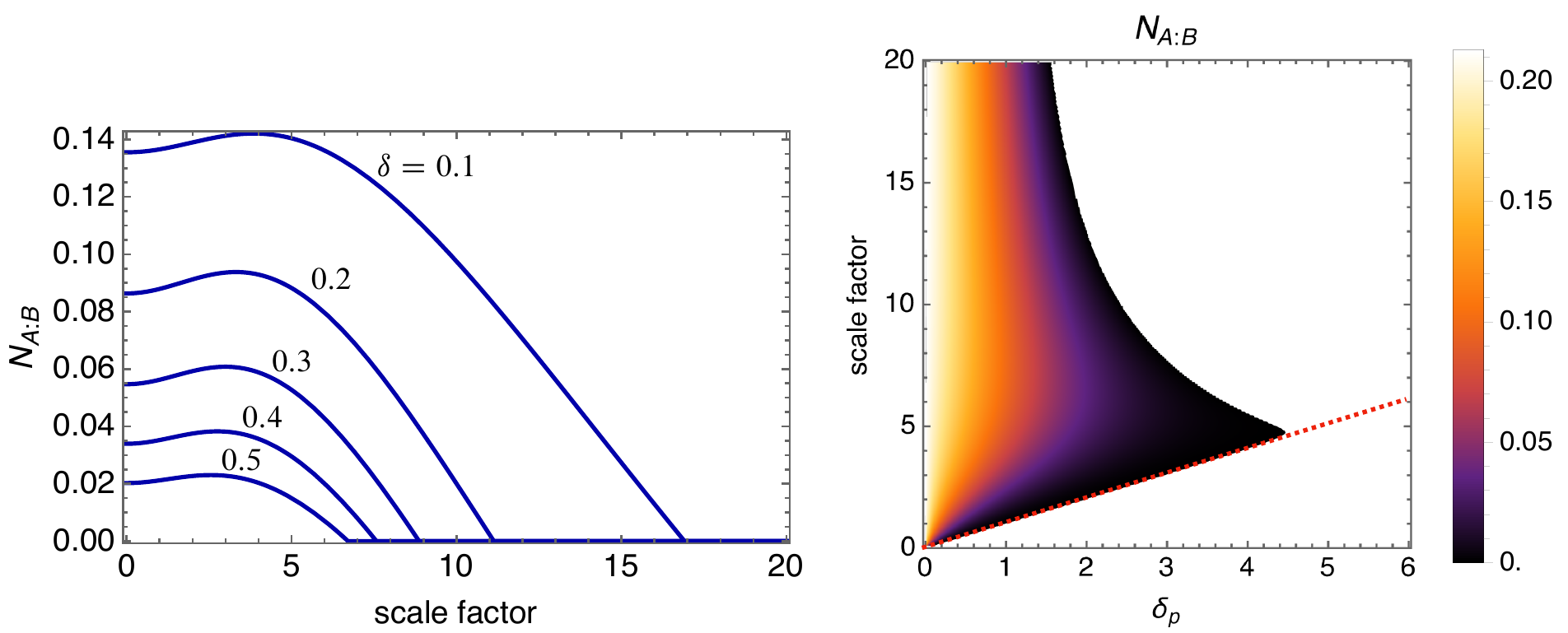}
  \caption{Behavior of negativity. Left panel: dependence on the scale
    factor with different values of the comoving size $\del$. Right
    panel: dependence on the physical size of the local region
    $\del_p=a_\text{sc}\del$ and the scale factor $a_\text{sc}$.  The
    dotted line is $\del_p=a_\text{sc}$ that represents the evolution
    of the total size of the universe. The modes A and B are initially
    entangled ($a_\text{sc}=0$) and become separable after the
    physical separation between them exceeds the Hubble horizon scale
    $\sim H^{-1}$ with $k_c/a_\text{sc}\ll k_0$ and the effect of the
    IR cutoff becomes negligible.}
   \label{fig:neg-D}
\end{figure}

We adopt the de Sitter expansion of the scale factor
$a_\text{sc}=-1/(H\eta),-\infty<\eta<0$, where $H$ is the Hubble
constant. Mode functions corresponding to the Bunch-Davies vacuum
state, which coincides with the Minkowski vacuum state in the short
wavelength limit, are given by
\begin{equation}
  f_k=\frac{1}{\sqrt{2|k|}}\left(1+\frac{1}{i|k|\eta}\right)e^{-i|k|\eta},\quad
    g_k=\sqrt{\frac{|k|}{2}}e^{-i|k|\eta}.
\end{equation}
Covariances of the field operators are calculated as
\begin{align}
  &M_{11}=\frac{1}{\pi}\int_0^\infty\frac{dk}{k}\left(1+\frac{1}{k^2\eta^2}\right)\cos
    k(x-y),\label{eq:covariance_11}\\
    &M_{22}=\frac{1}{\pi}\int_0^\infty dk\, k \cos
      k(x-y),\label{eq:covariance_22}\\
    &M_{12}=\frac{1}{\pi\eta}\int_0^\infty\frac{dk}{k}\cos k(x-y).\label{eq:covariance_12}
\end{align}
We choose the UV and IR cutoff in the window functions as
$k_c=\pi H/\del,k_0=\pi H$.
The IR cutoff represents the physical size of the whole universe
$a_\text{sc}H^{-1}$ and the UV cutoff represents the physical size of
the focusing spatial region
$\del\times a_\text{sc}H^{-1}, 0\le\del\le 1$.  Covariances \eqref{eq:c1}, \eqref{eq:c2}, \eqref{eq:c3} of the local modes  are
\begin{align}
  &c_{1}=\frac{1}{\pi H}\frac{\del}{1-\del}\int_\del^1\frac{dz}{z}
    \left(1+\frac{a_\text{sc}^2\, \del^2}{\pi^2z^2}\right)\cos\left(\frac{\pi z}{1+\del}\right),
  \\
  &c_{2}=\frac{\pi H}{\del(1-\del)}\int_\del^1 dz z\cos\left(\frac{\pi z}{1+\del}\right),\\
  &c_{3}=-\frac{1}{\pi
    H}\frac{a_\text{sc}H\del}{1-\del}\int_\del^1\frac{dz}{z}\cos\left(\frac{\pi z}{1+\del}\right).
\end{align}
The left panel of Fig. \ref{fig:neg-D} shows the evolution of
negativity of the bipartite mode AB, with a fixed comoving size
$\del$. The initial nonzero negativity evolves to be zero at  some
specific value of the scale factor.  The physical size of a local
region is characterized by $\del_p:=a_\text{sc}\del$. The right panel
of Fig. \ref{fig:neg-D} shows the plot of negativity as a function of
$(\del_p,a_\text{sc})$. For a fixed $\del\in(0,1)$, this figure shows
that entanglement (quantum correlation) between the two local modes A and
B is lost after the physical size of the comoving region exceeds the
Hubble horizon scale and the ``classical'' behavior of the quantum
field emerges
\cite{Nambu2008,Nambu2011,Nambu2009,Nambu2013,Matsumura2018}.

  
The disentanglement behavior in this figure can  be intuitively
understood as a result of the fact that ``thermal'' noise at the
Gibbons-Hawking temperature $T_H=H/(2\pi)$ associated with the
cosmological horizon destroys quantum correlations between A and
B. The rest of this paper aims to provide a more quantitative
understanding of the disentanglement phenomenon in terms of
entanglement monogamy. The bipartite state AB is usually mixed because
it is defined as a subsystem embedded in the total universe. As the
monogamy relation proposed by Camalet
\cite{Camalet2017,Camalet2017a,Camalet2018,Zhu2020,Zhu2021} suggests,
the amount of quantum correlation between A and B (i.e., internal
correlation) is affected by the amount of quantum correlation between
AB and its complement (i.e., external correlation). Therefore,
we look for the partner modes that purify the bipartite mode AB and
investigate the entanglement structure among them in the following
sections.

\section{Purification of local Gaussian modes  in quantum fields}\label{sec:3}
To understand the behavior of entanglement between
spatial local modes, we look for their partners, i.e., the modes that
purify them. In \cite{hotta2015partner}, a partner mode of a given
mode is calculated in specific examples, including a system with
Hawking radiation. A general partner formula that identifies a partner
mode for a single mode in any pure Gaussian state is proven in
\cite{Trevison2018b}. Generalizing these results, a systematic method
to identify any number of modes in a pure state has been developed in
\cite{Yamaguchi2020,Yamaguchi2020a,Yamaguchi2020b}. Such a subsystem
composed of modes in a pure state is called a quantum information
capsule (QIC). Although the QIC formula in
\cite{Yamaguchi2020,Yamaguchi2020a,Yamaguchi2020b} provides an
algebraic way to identify modes in a pure state, it cannot be directly
used for our purpose of analyzing the disentanglement structure of two
local modes AB \cite{Nambu2008,Nambu2011,Nambu2009,Nambu2013} from the
viewpoint of monogamy. In this section, we derive a more useful
formula to identify the partner modes that purify given two modes AB.

In Sec.~\ref{sec:QIC}, we briefly review the QIC argument with which modes in a pure state are identified. In Sec.~\ref{sec:partner_single_mode}, we derive a formula identifying the partner mode of a single mode, which reproduces the partner formula in \cite{Trevison2018b}. In Sec.~\ref{sec:partner_two_modes}, we generalize the partner formula to identify the partner of a two-mode system.



\subsection{QICs in Gaussian states}\label{sec:QIC}
The partners here are a special class of QIC. In
\cite{Yamaguchi2020,Yamaguchi2020a,Yamaguchi2020b}, it has been shown
that a linear map, denoted by $f_\psi$ for a pure Gaussian state
$\ket{\psi}$, plays a key role in identifying a QIC. We here briefly
review the results in
\cite{Yamaguchi2020,Yamaguchi2020a,Yamaguchi2020b}, including the
properties of $f_\psi$.

Let us first consider a system composed of $N$ harmonic oscillators,
which is assumed to be in a pure Gaussian state $\ket{\psi}$. The
canonical variables are defined by
$\hat{\bs{r}}=(\hat q_1,\hat p_1,\dots,\hat q_N,\hat p_N)^T$. For
simplicity, we assume that the first moments of the canonical
variables vanish, i.e., $\langle\hat{\bs{r}}\rangle=0$, where
$\langle\,\rangle$ denotes the expectation value in $\ket{\psi}$. The
covariance matrix with respect to these canonical variables is defined
by
$\bs{m}:=\langle\{\hat{\bs{r}},\hat{\bs{r}}^T\}\rangle$.\footnote{$[\hat{\bs{r}},\hat{\bs{r}}^T]$
  is the antisymmetric part of the operator $\hat{\bs{r}}\hat{\bs{r}}^T$
  and $\{\hat{\bs{r}},\hat{\bs{r}}^T\}$ is the symmetric part of the
  operator $\hat{\bs{r}}\hat{\bs{r}}^T$.} Because the total system is
assumed to be in a pure Gaussian state, all the symplectic eigenvalues
of $\bs{m}$ are one. That is, there exists a symplectic matrix
$\bs{S}$ such that
\begin{align}
    \bs{m}=\bs{S}\bs{S}^T,  \label{eq:pure-mode}
\end{align}
where $\bs{S}$ satisfies $\bs{S}^T\bs{\Omega}_N\bs{S}=\bs{\Omega}_N$ and $[\hat{\bs{r}},\hat{\bs{r}}^T]=i\bs{\Omega}_N$ for
\begin{equation}
  \bs{\Omega}_N:=\bigoplus_{i=1}^N\bs{J},\quad\bs{J}=
  \begin{bmatrix}
    0& 1\\ -1&0
  \end{bmatrix}.
\end{equation}
Note that the pure state condition \eqref{eq:pure-mode} is equivalent to the following
relation:
\begin{equation}
  \bs{m}\bs{\Omega}_N\bs{m}=\bs{\Omega}_N.\label{eq:purity_condition_discrete}
\end{equation}

If we introduce a new basis $\hat{\bs{r}}'$ for the canonical
variables by $\hat{\bs{r}}'=\bs{S}^{-1}\hat{\bs{r}}$, we get
\begin{equation}
  \bs{m}'=\langle\{\hat{\bs{r}}',\hat{\bs{r}}'^T\}\rangle=\bs{S}\bs{m}(\bs{S}^T)^{-1}=\mathbb{I}_{2N},\quad
  [\hat{\bs{r}}',\hat{\bs{r}}'^T]=i\bs{\Omega}_N. 
\end{equation}
The canonical variables defined by $\hat{\bs{r}}'$ specify $N$
uncorrelated modes, each of which is in a pure state. Now we define a linear map $f_\psi$ by
\begin{equation}
  f_\psi(\hat q_i')=\hat p_i',\quad   f_\psi(\hat p_i')=-\hat q_i',\label{eq:definition_f_diagonal_basis}
\end{equation}
or equivalently,
\begin{equation}
  f_\psi(\hat{\bs{r}}')=\bs{\Omega}_N\hat{\bs{r}}'.\label{eq:def_f_diagoal_vector}
\end{equation}
This map has the following properties:
\begin{align}
    [\hat{\bs{r}}',f_\psi(\hat{\bs{r}}')^T]&=i\mathbb{I}_{2N},\label{eq:ccr_map_f}\\
    \langle\{\hat{\bs{r}}',\hat{\bs{r}}'^T\}\rangle&=\mathbb{I}_{2N},\quad \langle\{\hat{\bs{r}}',f_\psi(\hat{\bs{r}}')^T\}\rangle=0,\quad \langle\{f_\psi(\hat{\bs{r}}'),f_\psi(\hat{\bs{r}}')^T\}\rangle=\mathbb{I}_{2N}.\label{eq:cov_map_f}
\end{align}
On the one hand, Eq.~\eqref{eq:ccr_map_f} means that
$[\hat{r_j}',f_\psi(\hat{r_j}')]=i$; i.e.,
$(\hat{r_j}',f_\psi(\hat{r_j}'))$ defines a mode for any
$j=1,\dots, 2N$. On the other hand, Eq.~\eqref{eq:cov_map_f} implies
that the mode defined by $(\hat{r_j}',f_\psi(\hat{r_j}'))$ is in a
pure state. Of course, these are straightforward consequences of the
definition in Eq.~\eqref{eq:definition_f_diagonal_basis}.

Let us generalize this observation.  Consider an operator $\hat{O}$
given by a linear combination of canonical variables. We first rescale
this operator as $\hat{O}\to \hat{O}/\sqrt{2\langle\hat{O}^2\rangle}$
so that $\langle\hat{O}^2\rangle=1/2$. Expanding $\hat{O}$ in the
basis $\hat{\bs{r}}'$ as
\begin{align}
    \hat{O}:=\sum_{i=1}^N w_i'\hat{r}_i'=\bs{w}^{\prime T}\hat{\bs{r}}',
\end{align}
where $\bs{w}^{\prime T}=(w_1',w_2',\dots,w_{2N}')$, the
normalization condition is equivalent to
\begin{align}
    \bs{w}^{\prime T}\bs{w}'=1.\label{eq:coefficient_vector_normalization}
\end{align}
Introducing operators $(\hat{Q},\hat{P}):=(\hat{O},f_\psi(\hat{O}))$
for $\hat{O}$ with this normalization, from Eqs.~\eqref{eq:ccr_map_f},
\eqref{eq:cov_map_f} and \eqref{eq:coefficient_vector_normalization},
we find
\begin{align}
    [\hat{Q},\hat{P}]&= i\label{eq:ccr_1qic}\\
    \langle\{\hat{Q},\hat{Q}\}\rangle&=1,\quad \langle\{\hat{Q},\hat{P}\}\rangle=\langle\{\hat{P},\hat{Q}\}\rangle=0,\quad \langle\{\hat{P},\hat{P}\}\rangle=1.\label{eq:cov_1qic}
\end{align}
Equation~\eqref{eq:ccr_1qic} means that $(\hat{Q},\hat{P})$ satisfies
the canonical commutation relation and hence defines a
mode. Further, Eq.~\eqref{eq:cov_1qic} implies that the covariance
matrix for this mode is equal to the $2\times 2$ identity matrix,
implying that it is in a pure state.

In a series of studies
\cite{yamaguchi2019quantum,Yamaguchi2020,Yamaguchi2020a,Yamaguchi2020b}
on the carriers of information, the smallest subsystem in a pure state
that carries the whole encoded information is termed a QIC. The
encoded information can be fully retrieved by extracting a QIC from
the system. Equations~\eqref{eq:ccr_1qic} and \eqref{eq:cov_1qic}
imply that the mode defined by
$(\hat{Q},\hat{P}):=(\hat{O},f_\psi(\hat{O}))$ is the QIC when the 
encoding operation is generated by $\hat{O}$. In a more general case
where the encoding operation is generated by $\{\hat{O}_i\}_{i=1}^n$,
where each of which is assumed to be a linear combination of canonical
variables, it is proven \cite{Yamaguchi2020} that the QIC is given by
a subsystem composed of (at most) $n$ modes, which is algebraically
defined by
\begin{align}
   \{(\hat{O}_i,f_\psi(\hat{O}_i)\}_{i=1}^n.
\end{align}
It is shown that operators $\{(\hat{Q}_i,\hat{P}_i)\}_i$
defined in Eq.~(73) in \cite{Yamaguchi2020} satisfy
\begin{align}
    [\hat{Q}_j,\hat{P}_k]&= i\delta_{jk},\quad [\hat{Q}_j,\hat{Q}_k]=[\hat{P}_j,\hat{P}_k]=0\label{eq:ccr_nqic}\\
    \langle\{\hat{Q}_j,\hat{Q}_k\}\rangle&=\delta_{jk},\quad \langle\{\hat{Q}_j,\hat{P}_k\}\rangle=\langle\{\hat{P}_j,\hat{Q}_k\}\rangle=0,\quad \langle\{\hat{P}_j,\hat{P}_k\}\rangle=\delta_{jk},\label{eq:cov_nqic}
\end{align}
which generalizes Eqs.~\eqref{eq:ccr_nqic} and \eqref{eq:cov_nqic}. It
implies that the mode characterized by $\{(\hat{Q}_i,\hat{P}_i)\}_i$
is in a pure state when the total system is in $\ket{\psi}$, and
therefore, they are the QIC as the encoded information is
carried by them.

Although the subsystem composed of $n$ modes playing the role of QIC
is uniquely determined, there are several ways to decompose it into
$n$ independent modes. From Eq.~\eqref{eq:cov_nqic}, the Gaussian
state $\ket{\psi}$ of the total system is decomposed into
\begin{align}
    \ket{\psi}=\ket{\psi'}_{12\cdots n} \otimes \ket{\psi''}_{\overline{12 \cdots n}},\quad \ket{\psi'}_{12\cdots n} :=\bigotimes_{j=1}^n\ket{0}_j
\end{align}
where $\ket{0}_j$ denotes the ``vacuum'' state for the $j$th mode
annihilated by $\hat{a}_j:=(\hat{Q}_j+i\hat{P}_j)/\sqrt{2} $ and
$\ket{\psi''}_{\overline{12 \cdots n}}$ denotes a pure state for the
complementary system. Since each of the $n$ modes is in a pure state in
this decomposition, the analysis of the entanglement structure
  is not straightforward. In the following subsections, we introduce
another decomposition that is more useful in analyzing the
entanglement structure among partners.

For later convenience, we summarize here the properties of
$f_\psi$. From the definition in Eq.~\eqref{eq:def_f_diagoal_vector},
$f_\psi$ maps the original basis $\hat{\bs{r}}$ to
\begin{equation}
  f_\psi(\hat{\bs{r}})=f_\psi(\bs{S}\hat{\bs{r}}')=\bs{S}\bs{\Omega}_N\bs{S}^{-1}\hat{\bs{r}}=\bs{m}\bs{\Omega}_N\hat{\bs{r}},\label{eq:def_map_f_in_orignal_basis}
\end{equation}
where we have used
$\bs{S}^{-1}=-\bs{\Omega}_{N}\bs{S}^T\bs{\Omega}_N$. From
Eqs.~\eqref{eq:ccr_map_f} and \eqref{eq:cov_map_f}, for any operators
$\hat O$ and $ \hat O'$ given by linear combinations of canonical
operators by \eqref{eq:local-op} \cite{Yamaguchi2020a,Yamaguchi2020},
it can be directly checked that
\begin{align}
  &\langle\hat O\rangle=0,\quad\langle f_\psi(\hat O)\rangle=0,\quad
    f_\psi(f_\psi(\hat O))=-\hat O\label{eq:map0}\\
    & [\hat O, f_\psi(\hat
    O')]=i\langle\{\hat O,\hat O'\}\rangle, \quad
    \langle\{\hat O,f_\psi(\hat O')\}\rangle=i[\hat O,\hat O'], \label{eq:map1}\\
  &[f_\psi(\hat O), f_\psi(\hat O')]=[\hat O,\hat O'],\quad
    \langle \{f_\psi(\hat O),f_\psi(\hat O')\}\rangle=\langle\{\hat O,\hat
    O'\}\rangle \label{eq:map2}
\end{align}
hold. 

So far, we have reviewed the properties of map $f_\psi$ in a harmonic
oscillator system. The analyses can  readily be extended to a scalar
field by the following procedure. For a scalar field $\hat{\varphi}$
and its conjugate momentum $\hat{\pi}$ at a fixed time $t$, we denote
\begin{align}
    \hat{\bs{R}}(x):=
    \begin{bmatrix}\hat\varphi(t,x)\\\hat\pi(t,x)
    \end{bmatrix}.
\end{align}
Here, for notational simplicity, we omit the time variable $t$ on the
left-hand side. The equal-time commutation relations in
Eq.~\eqref{eq:equal_time_ccr} are written as
\begin{equation}
  [\hat{\bs{R}}(x),\hat{\bs{R}}^T\!(y)]=i
  \begin{bmatrix}
    0&\del(x-y) \\ -\del(x-y) &0
  \end{bmatrix}=i\bs{J}\del(x-y),
\end{equation}
where $\bs{J}=\begin{bmatrix}0&1\\-1&0\end{bmatrix}$, while the
covariance of the field operators in the state $\ket{\psi}$ are denoted by
\begin{equation}
  \bs{M}(x,y):=\langle\{\hat{\bs{R}}(x),\hat{\bs{R}}^T\!
  (y)\}\rangle=
  \begin{bmatrix}
      M_{11}(x,y)&M_{12}(x,y)\\
      M_{21}(x,y)&M_{22}(x,y)
  \end{bmatrix}=
  \begin{bmatrix}
    \langle\{\hat\varphi(x),\hat\varphi(y)\}\rangle &           \langle\{\hat\varphi(x),\hat\pi(y)\}\rangle \\
    \langle\{\hat\pi(x),\hat\varphi(y)\}\rangle &\langle\{\hat\pi(x),\hat\pi(y)\}\rangle 
  \end{bmatrix}
  .
\end{equation}
We introduce an operator
\begin{equation}
  \hat O:=\int
  dx\left[w_1(x)\hat\varphi(x)+w_2(x)\hat\pi(x)\right]
  =\int dx \bs{W}^T\!(x)\hat{\bs{R}}(x) ,
  \label{eq:local-op}
\end{equation}
where
\begin{equation}
    \bs{W}(x)= \begin{bmatrix}w_1(x)\\ w_2(x)
                        \end{bmatrix}
\end{equation}
denotes weighting functions. In the analogy with Eq.~\eqref{eq:def_map_f_in_orignal_basis}, we define \cite{Yamaguchi2020} a map $f_\psi$ by
\begin{align}
    f_\psi(\hat{O})&:=\int dxdy  dz\bs{W}^T(x)\bs{M}(x,y)\bs{J}\delta(y-z)\hat{\bs{R}}(z)\notag\\
    &=\int dx \bs{W}_{\!\!\!f_\psi(\hat{O})}^T(x)\hat{\bs{R}}(x),
    \label{eq:f}
\end{align}
where $\bs{W}_{\!\!\!f_\psi(\hat{O})}$ is the window function defining $f_\psi(\hat O)$, given by
\begin{equation}
  \bs{W}_{\!\!\!f_\psi(\hat{O})}(x):=-\int dy\bs{J}\bs{M}(x,y)\bs{W}(y).
  \label{eq:wf}
\end{equation}
When the covariance of the field operator satisfies a purity condition
\begin{align}
    \int dydz \bs{M}(x,y)\bs{J}\delta(y-z)\bs{M}(z,w)=\bs{J}\delta(x-w),\label{eq:purity_condition_field}
\end{align}
which corresponds to Eq.~\eqref{eq:purity_condition_discrete}, it is
shown \cite{Yamaguchi2020a,Yamaguchi2020} that the map $f_\psi$
satisfies all the properties in Eqs.~\eqref{eq:map0}, \eqref{eq:map1}
and \eqref{eq:map2}. Therefore, for operators $\{O_i\}_{i=1}^n$ given
by linear combinations of field operators, the set of operators
\begin{align}
   \{(\hat{O}_i,f_\psi(\hat{O}_i)\}_{i=1}^n
\end{align}
defines (at most) $n$ modes in a pure state, provided that the field
is in a pure Gaussian state $\ket{\psi}$. Note that
Eq.~\eqref{eq:purity_condition_field} can be explicitly confirmed for
the covariances given in Eqs.~\eqref{eq:covariance_11},
\eqref{eq:covariance_22} and \eqref{eq:covariance_12}.

\begin{figure}[H]
  \centering
  \includegraphics[width=0.4\linewidth]{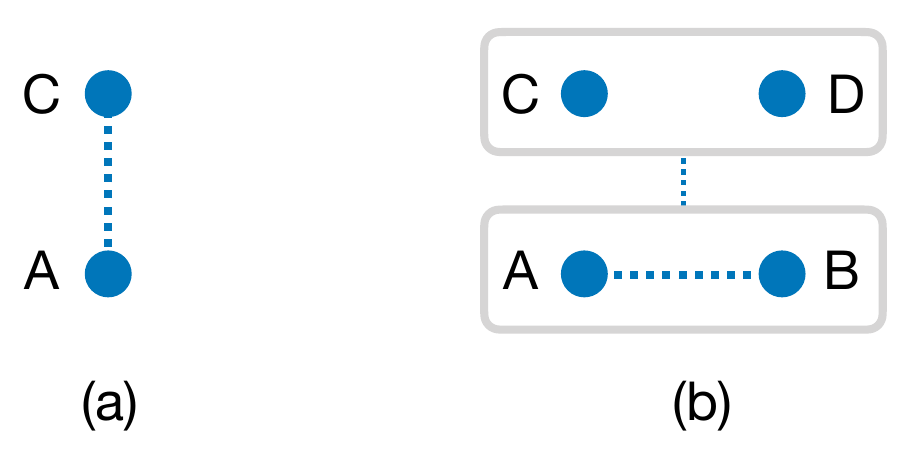}
  \caption{(a) Purification of a single mode A. The mode C is a partner
    of A. (b) Purification of
    two modes AB. The modes C and D are partners of the bipartite
    system AB.}
  \label{fig:purification}
\end{figure}

Based on these results, we derive the partner formula for a single
mode in Sec.~\ref{sec:partner_single_mode}, which reproduces the
formula in \cite{Trevison2018b}. We further generalize it for the
partner formula for two modes in Sec.~\ref{sec:partner_two_modes},
which we shall use to analyze entanglement monogamy among local modes
in a field. See Fig.~\ref{fig:purification} for the schematic picture
of these setups.
\subsection{Purification of a single Gaussian mode}\label{sec:partner_single_mode}
As a practical application of the map $f_\psi$, we look for a partner
mode that purifies a given single mode A [Fig. \ref{fig:purification}
(a)].  In particular, we apply the partner formula for a local mode
$\hat{\bs{\xi}}_\text{A}=(\hat q_\text{A},\hat p_\text{A})^T$  at a
spatial point ${\bs{x}}_\text{A}$ defined in the previous section.
As we have seen in Sec.~\ref{sec:QIC}, four operators
\begin{equation}
  \hat q_\text{A}, \hat p_\text{A}, f_\psi(\hat q_\text{A}), f_\psi(\hat p_\text{A}),\label{eq:four_ops_1partner}
\end{equation}
define a two-mode system that is in a pure state, provided that the field is in a pure Gaussian state. To identify the partner mode of
$\hat{\bs{\xi}}_\text{A}$, we here construct a mode generated by the operators in Eq.~\eqref{eq:four_ops_1partner}, which is orthonormal to the mode A. The
covariance matrix for  $\hat{\bs{\xi}}_\text{A}$ is
\begin{equation}
\bs{m}_\text{A}:=\expval{\{\hat{\bs{\xi}}_\text{A},\hat{\bs{\xi}}^T_\text{A}
    \}}=
\begin{bmatrix}
  2\expval{\hat q_\text{A}^2}& \expval{\hat q_\text{A}\hat p_\text{A}+\hat p_\text{A}\hat q_\text{A}}\\
  \expval{\hat q_\text{A}\hat p_\text{A}+\hat p_\text{A}\hat q_\text{A}}& 2\expval{\hat p_\text{A}^2}
\end{bmatrix}
=
\begin{bmatrix}
  c_1&c_3\\c_3&c_2
\end{bmatrix}
.
\end{equation}
Commutators and covariances between these operators are given by
\begin{equation}
  [\hat{\bs{\xi}}_\text{A},\hat{\bs{\xi}}^T_\text{A}]=[f_\psi(\hat{\bs{\xi}}_\text{A}),f_\psi(\hat{\bs{\xi}}^T_\text{A})]=i\bs{J},\quad
  [\hat{\bs{\xi}}_\text{A},f_\psi(\hat{\bs{\xi}}^T_\text{A})]=i\bs{m}_\text{A},
\end{equation}
and
\begin{equation}
  \expval{\{\hat{\bs{\xi}}_\text{A},\hat{\bs{\xi}}^T_\text{A}
    \}}=
  \expval{\{f_\psi(\hat{\bs{\xi}}_\text{A}),
    f_\psi(\hat{\bs{\xi}}^T_\text{A}) \}}=\bs{m}_\text{A},
  \quad \expval{\{\hat{\bs{\xi}}_\text{A},
    f_\psi(\hat{\bs{\xi}}^T_\text{A}) \}}=-\bs{J}.
\end{equation}
To extract a mode orthogonal to the original mode
$\hat{\bs{\xi}}_\text{A}$ from $f_\psi(\hat{\bs{\xi}}_\text{A})$, we
define operators
\begin{equation}
  \hat{\bs{\zeta}}:=f_\psi(\hat{\bs{\xi}}_\text{A})-\bs{m}_\text{A}\bs{J}\hat{\bs{\xi}}_\text{A}.
\end{equation}
They indeed commute with $\hat{\bs{\xi}}_\text{A}$ as
\begin{align}
  [\hat{\bs{\zeta}},\hat{\bs{\xi}}^T_\text{A}]
  &=[f_\psi(\hat{\bs{\xi}}_\text{A}),\hat{\bs{\xi}}^T_\text{A}]-[\bs{m}_\text{A}
    \bs{J}\hat{\bs{\xi}}_\text{A},\hat{\bs{\xi}}^T_\text{A}]
    \notag\\
  &=-\bs{m}_\text{A}-\bs{m}_\text{A}\bs{J}[\hat{\bs{\xi}}_\text{A},\hat{\bs{\xi}}^T_\text{A}]\notag\\
  &=0.
\end{align} 
Therefore, they define a mode
orthogonal to $\hat{\bs{\xi}}_\text{A}$.

The commutator of $\hat{\bs{\zeta}}$ is calculated as
\begin{align}
  [\hat{\bs{\zeta}},\hat{\bs{\zeta}}^T]&=i(\bs{J}-\bs{m}_\text{A}\bs{J}\bs{m}_\text{A}).
\end{align}
If the mode A is in a pure state, it holds
$\bs{m}_\text{A}\bs{J}\bs{m}_\text{A}=\bs{J}$ which corresponds to the
relation of the density operator
$\hat \rho^2_\text{A}=\hat\rho_\text{A}$, implying that the commutator
of $\hat{\bs{\zeta}}$ is trivial. In this case, because
$\hat{\bs{\xi}}_\text{A}$ is in a pure state, its partner mode does
not exist. If the mode A is not pure, its partner is characterized by
$\hat{\bs{\zeta}}$.
We normalize $\hat{\bs{\zeta}}$ to make it a canonical pair of
operators. For this purpose, we introduce $\hat{\bs{\xi}}_\text{C}$ by
$\hat{\bs{\zeta}}=\bs{A}\,\hat{\bs{\xi}}_\text{C}$ with a matrix $\bs{A}$ so that 
$[\hat{\bs{\xi}}_\text{C}, \hat{\bs{\xi}}_\text{C}^T]=i\bs{J}$ holds. The
condition on the matrix $\bs{A}$ is given by
\begin{equation}
  \bs{AJA}^T=\bs{J}-\bs{m}_\text{A}\bs{J}\bs{m}_\text{A}.\label{eq:matrix_a_constraint}
\end{equation}
To obtain $\bs{A}$, we consider the standard form of the covariance
matrix $\bs{m}_\text{A}$:
\begin{equation}
  \bs{m}_\text{A}=  \bs{S}\begin{bmatrix}
    a&0\\0&a
  \end{bmatrix}\bs{S}^T
  =a\bs{S}\bs{S}^T,\quad \bs{SJS^T}=\bs{J},
\end{equation}
where $\bs{S}$ represents a symplectic transformation to diagonalize
$\bs{m}_\text{A}$ and $a$ is the symplectic eigenvalue of
$\bs{m}_\text{A}$. 

Although the partner mode C itself is unique,
the matrix $\bs{S}$ satisfying Eq.~\eqref{eq:matrix_a_constraint} is
not uniquely determined because of the remaining freedom in fixing a
canonical set of operators for the mode C. In other words, when
$\bs{A}$ satisfies Eq.~\eqref{eq:matrix_a_constraint}, so does
$\bs{A}':=\bs{S}'\bs{A}$, where $\bs{S}'$ is an arbitrary $2\times 2$
symplectic matrix. Since Eq.~\eqref{eq:matrix_a_constraint} is recast
into $\bs{AJA}^T=(1-a^2)\bs{J}$, we can choose
\begin{equation}
\bs{A}
  =\sqrt{a^2-1}\,\bs{X}  ,
\end{equation}
or equivalently 
\begin{align}
  \bs{A}^{-1}=\frac{1}{\sqrt{a^2-1}}\bs{X},\quad
\end{align}
where $\bs{X}=
\begin{bmatrix}
    0&1\\1&0
  \end{bmatrix}$.

In summary, the partner mode C of mode A is obtained as the following formula:
\begin{align}
  \hat{\bs{\xi}}_\text{C}&=\bs{A}^{-1}\hat{\bs{\zeta}}
  =\frac{1}{\sqrt{a^2-1}}\bs{X}
                           (f_\psi(\hat{\bs{\xi}}_\text{A})-\bs{m}_\text{A}
                           \bs{J}\hat{\bs{\xi}}_\text{A}) \notag \\
 &=\frac{1}{\sqrt{a^2-1}}
  \begin{bmatrix}
    c_2\hat q_\text{A}-c_3\hat p_\text{A}+f_\psi(\hat p_\text{A})\\
    c_3\hat q_\text{A}-c_1\hat p_\text{A}+f_\psi(\hat
    q_\text{A})
  \end{bmatrix}
  =:
  \begin{bmatrix}
    \hat q_\text{C}\\ \hat p_\text{C}
  \end{bmatrix},
  \label{eq:partner1}
\end{align}
which is equivalent to the partner formula for a single mode \cite{Trevison2018b} in a Gaussian state.
Covariances of operators 
$(\hat{\bs{\xi}}_\text{A}, \hat{\bs{\xi}}_\text{C})$ are given by
\begin{align}
 \expval{\{\hat{\bs{\xi}}_\text{A},\hat{\bs{\xi}}_\text{A}^T\}}
  &=\bs{m}_\text{A},\\
  \expval{\{\hat{\bs{\xi}}_\text{C},\hat{\bs{\xi}}_\text{A}^T\}}
  &=\bs{A}^{-1}(\bs{J}-\bs{m}_\text{A}\bs{J}\bs{m}_\text{A})=\sqrt{a^2-1}\bs{Z},\\
  \expval{\{\hat{\bs{\xi}}_\text{C},\hat{\bs{\xi}}_\text{C}^T\}}
  &=-\bs{A}^{-1}(\bs{m}_\text{A}+\bs{m}_\text{A}\bs{J}\bs{m}_\text{A}\bs{J}\bs{m}_\text{A})
    (\bs{A}^{-1})^T
    =\bs{X}\bs{m}_\text{A}\bs{X},
\end{align}
where $\bs{Z}=
  \begin{bmatrix}
    1&0\\0&-1
  \end{bmatrix}$.
In a matrix form, they are summarized as
\begin{equation}
  \bs{m}_\text{AC}:=\expval{\{\hat{\bs{\xi}}_{\text{AC}},\hat{\bs{\xi}}_{\text{AC}}^T\}}=
  \begin{bmatrix}
    \bs{m}_\text{A}&\sqrt{a^2-1}\bs{Z}\\\sqrt{a^2-1}\bs{Z}& \bs{X}\bs{m}_\text{A}\bs{X}
  \end{bmatrix},\quad 
  \hat{\bs{\xi}}_{\text{AC}}:=
  \begin{bmatrix}
      \hat{\bs{\xi}}_{\text{A}}\\
      \hat{\bs{\xi}}_{\text{C}}
  \end{bmatrix}.
  \label{eq:MAC}
\end{equation}
One can explicitly confirm that this covariance matrix satisfies the following purity condition of the state AC:
\begin{equation}
  \bs{m}_{\text{AC}}\,\bs{\Omega}_2\,\bs{m}_{\text{AC}}=\bs{\Omega}_2,\quad
  \bs{\Omega}_2=\bigoplus_{i=1}^2\bs{J},
\end{equation}
implying that it represents a pure two-mode squeezed state characterizing the pair of partners AC. The total state is decomposed as
\begin{equation}
  \ket{\psi}=\ket{\psi'}_\text{AC}\otimes\ket{\psi''}_{\overline{\text{AC}}},
\end{equation}
where $\ket{\psi'}_\text{AC}$ is the pure Gaussian state defined by
$\bs{m}_\text{AC}$ and has no correlation with its complement system $\overline{\text{AC}}$ in another pure state $\ket{\psi''}_{\overline{\text{AC}}}$.

\begin{figure}[H] 
  \centering
  \includegraphics[width=1.0\linewidth]{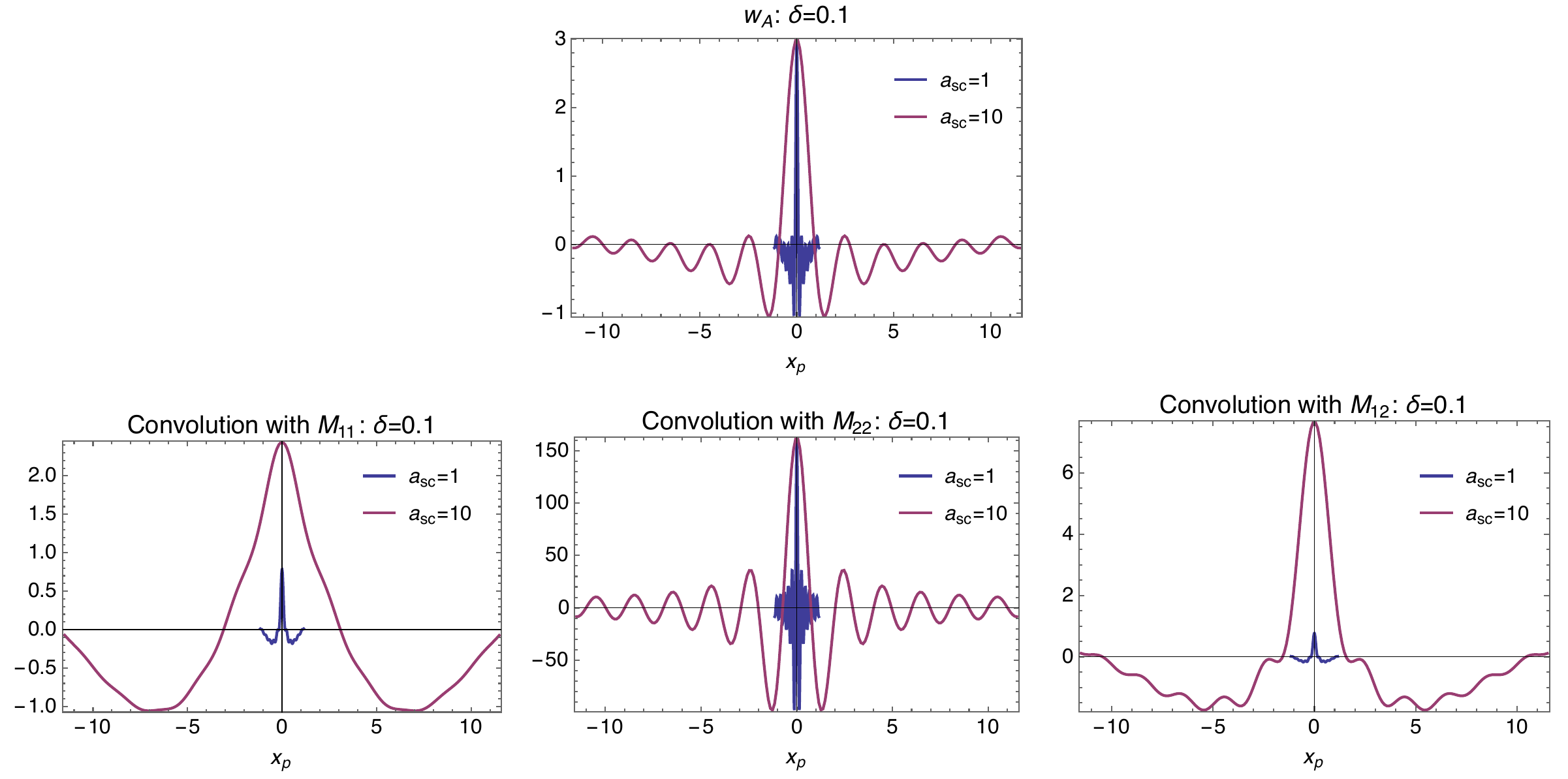}
  \caption{Upper panel: the window function $w_\text{A}$ that
    represents the spatial profile of mode A at
    $a_\text{sc}=1,10$. We set $x_\text{A}=0$. $x_p=a_\text{sc}\,x$ denotes the
    physical coordinate. Owing to cosmic expansion, the width of the
    profile increases as $ a_\text{sc}\del$. Lower panels: convolution
    of $w_\text{A}$ with covariances of the field operators. These functions
    represent spatial profiles of the partner mode C.}
  \label{fig:partner-profile}
\end{figure}

\begin{figure}[H] 
  \centering
   \includegraphics[width=0.8\linewidth]{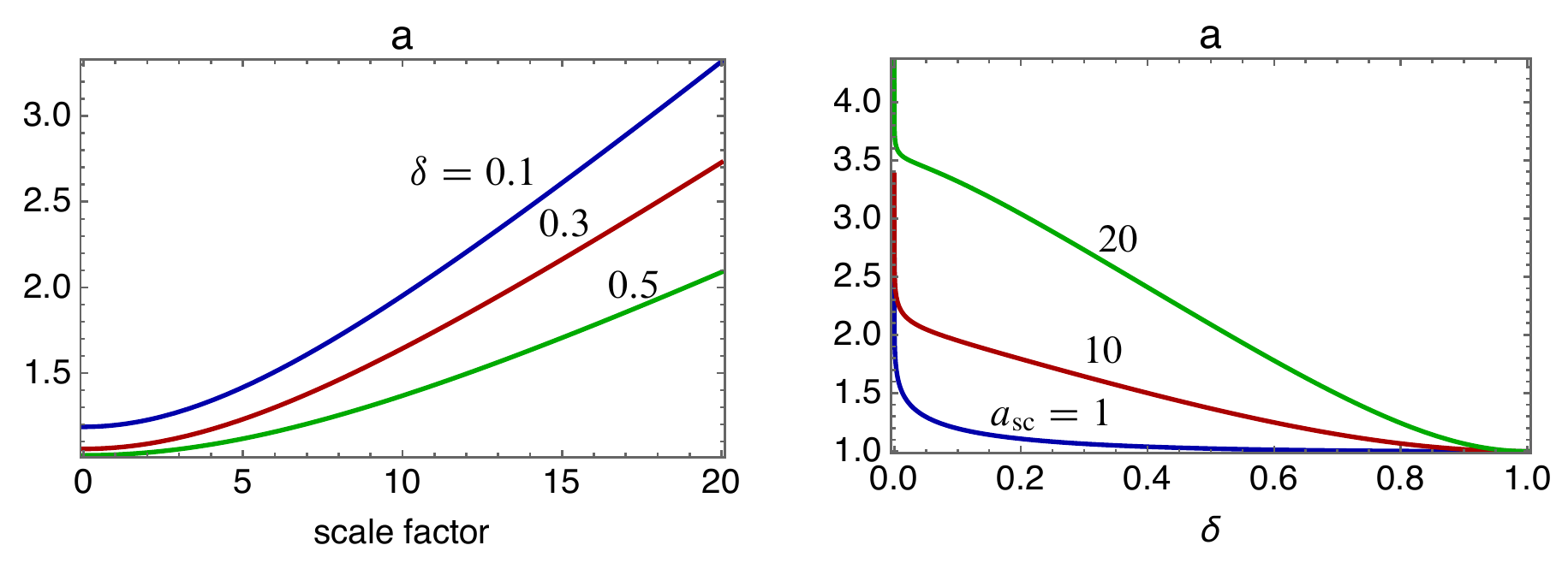}
  \caption{Behavior of the symplectic eigenvalue $a$. Left panel:
    dependence on the scale factor. The symplectic eigenvalue $a$ is
    an increasing function of the scale factor and entanglement between
    AC grows as the universe expands. Right panel: dependence on the normalized comoving
    size $\del$ of the region A.}
  \label{fig:single-region}
\end{figure}

\paragraph{Spatial profile of partner mode.}
Using Eq. \eqref{eq:wf}, the spatial profiles of the
partner mode can be visualized. As a window function of local mode A, we
adopt
$\bs{W}_{\!\!q_\text{A}}=w_\text{A}(x)(1,0)^T$ and $\bs{W}_{\!\!p_\text{A}}=w_\text{A}(x)(0,1)^T$, which we have used to 
introduce local modes from the scalar field in \eqref{eq:local-mode1} and
\eqref{eq:local-mode2}. From Eq.~\eqref{eq:wf}, we get 
\begin{align}
  &\bs{W}_{\!\!\!f_\psi(q_\text{A})}(x)=-\int dy
  \begin{bmatrix}
    M_{12}(x,y)&M_{22}(x,y) \\ -M_{11}(x,y)&-M_{12}(x,y)
  \end{bmatrix}
    \begin{bmatrix}
      w_\text{A}(y)\\ 0
    \end{bmatrix}
    =\int dy
    \begin{bmatrix}
      -M_{12}(x,y) \\M_{11}(x,y)
    \end{bmatrix}
    w_\text{A}(y), \\
  &\bs{W}_{\!\!\!f_\psi(p_\text{A})}(x)=-\int dy
  \begin{bmatrix}
    M_{12}(x,y)&M_{22}(x,y) \\ -M_{11}(x,y)&-M_{12}(x,y)
  \end{bmatrix}
    \begin{bmatrix}
      0\\ w_\text{A}(y)
    \end{bmatrix}
    =\int dy
    \begin{bmatrix}
      -M_{22}(x,y) \\M_{12}(x,y)
    \end{bmatrix}
    w_\text{A}(y), 
\end{align}
and
\begin{align}
  &f_\psi(\hat{q}_\text{A})=\int dx \bs{W}_{\!\!\!f_\psi(q_\text{A})}^T(x)\hat{\bs{R}}(x)=\int dx
    dy\left[-\hat\varphi(x)M_{12}(x,y)w_\text{A}(y)+\hat\pi(x)M_{11}(x,y)w_\text{A}(y)
    \right], \label{eq:partner-f1}\\
    &f_\psi(\hat{p}_\text{A})=\int dx \bs{W}_{\!\!\!f_\psi(p_\text{A})}^T(x)\hat{\bs{R}}(x)=\int dx
    dy\left[-\hat\varphi(x)M_{22}(x,y)w_\text{A}(y)+\hat\pi(x)M_{12}(x,y)w_\text{A}(y)
    \right]. \label{eq:partner-f2}
\end{align}
In other words, the window functions of
$f_\psi(\hat{\bs{\xi}}_\text{A})$ are expressed by convolutions of the
window function $w_\text{A}$ with the covariance matrix $M_{ij}$ of
the field operators, given by
\begin{align}
  \int dy
    M_{11}(x,y)w_\text{A}(y)&=\frac{4}{\sqrt{\pi(k_c-k_0)}}\int_{k_0}^{k_c}dk|f_k|^2\cos
                     k(x-x_\text{A}) \notag\\
  &=\frac{2}{\pi\sqrt{H}}\sqrt{\frac{\del}{1-\del}}\int_\del^1\frac{dz}{z}\left(
    1+\frac{a_\text{sc}^2\,\del^2}{\pi^2z^2}\right)\cos\left(z\frac{\pi H
    (x-x_\text{A})}{\del}\right),\\
    \int dy
      M_{22}(x,y)w_\text{A}(y)&=\frac{4}{\sqrt{\pi(k_c-k_0)}}\int_{k_0}^{k_c}dk|g_k|^2\cos
                       k(x-x_\text{A}) \notag \\
  &=\frac{2}{\pi\sqrt{H}}\sqrt{\frac{\del}{1-\del}}\left(\frac{\pi
      H}{\del}\right)^2\int_\del^1 dz z\cos\left(z\frac{\pi H
    (x-x_\text{A})}{\del}\right),\\
    \int dy
      M_{12}(x,y)w_\text{A}(y)&=\frac{2}{\sqrt{\pi(k_c-k_0)}}\int_{k_0}^{k_c}dk\,
                       i(f_kg_k^*-f_k^*g_k)\cos k(x-x_\text{A}) \notag \\
      &=-\frac{2}{\pi\sqrt{H}}\sqrt{\frac{\del}{1-\del}}\,(a_\text{sc}H)\int_\del^1\frac{dz}{z}
    \cos\left(z\frac{\pi H (x-x_\text{A})}{\del}\right).
\end{align}
From these equations, we obtain spatial profiles of the partner mode
C.  The upper panel of Fig. \ref{fig:partner-profile} shows the window
function of the mode A with $\del=0.1$ as a function of the physical
coordinate $x_p:=a_\text{sc}\,x$ (we set $H=1$). Because of cosmic
expansion, the width of the window function (spatial size of the local
mode A) increases from $0.1H^{-1}$ to $H^{-1}$. The lower panels are
the convolution of $w$ with the covariance of field operators, which
appear in the partner formulas \eqref{eq:partner-f1} and
\eqref{eq:partner-f2}. As we can observe from the behavior of
convolutions with $M_{11},M_{12}$, amplitudes of these functions
become larger as the universe expands and typical wavelengths become
$\sim 10H^{-1}$ which is far larger than the width of $w_\text{A}$;
this behavior of the partner's window functions implies that the
information of the original mode A shared with its partner C
delocalizes and extends over the superhorizon scale. These facts
provide the following intuitive understanding of the mechanism of
disentanglement between local modes: Because the partner of a local
mode A is spread over the superhorizon scale for a large scale
factor, it is slightly different from another local mode B, implying that
modes AB cannot share much entanglement. This observation becomes more
quantitative in Sec.~\ref{sec:4}, where we analyze the
disentanglement from the viewpoint of entanglement monogamy.
\paragraph{Negativity between AC.}
The standard form of the covariance matrix for the mode A is
\begin{equation}
  \bs{m}_\text{A}=
  \begin{bmatrix}
    a & 0 \\ 0&a
  \end{bmatrix},
\end{equation}
where $a$ is the symplectic eigenvalue of $\bs{m}_\text{A}$. 
As shown in Eq.~\eqref{eq:MAC}, the covariance matrix of A and its partner C is given by
\begin{equation}
 \bs{m}_\text{AC}=
  \begin{bmatrix}
    a&0&\sqrt{a^2-1} &0 \\
    0&a&0&-\sqrt{a^2-1}\\
    \sqrt{a^2-1}&0&a&0\\
    0&-\sqrt{a^2-1}&0&a
  \end{bmatrix}.
\end{equation}
The smaller symplectic eigenvalue of its partial transposition of $\bs{m}_\text{AC}$ is calculated as
\begin{equation}
  \widetilde{\nu}_{2}=a-\sqrt{a^2-1}=\frac{1}{a+\sqrt{a^2-1}}\le 1.
\end{equation}
Thus the negativity between the modes A and C is given by 
\begin{equation}
  N_\text{A:C}=\frac{1}{2}(a+\sqrt{a^2-1}-1)\ge 0.
\end{equation}
For $a>1$, the bipartite state AC is entangled. Figure
\ref{fig:single-region} shows the behavior of the symplectic eigenvalue
$a=\sqrt{a_1a_2-a_3^2}$ as functions of the scale factor and $\delta$,
where $a_1,a_2,a_3$ are components of the covariance matrix
\eqref{eq:mAB}. The left panel of Fig. \ref{fig:single-region} shows
that for a fixed value of $\del$, the symplectic eigenvalue $a$ and
the negativity between AC increase as the scale factor increases. The
right panel of Fig. \ref{fig:single-region} shows that for a fixed
value of scale factor $a_\text{sc}$, the symplectic eigenvalue $a$
increases as the size $\del$ of the region A decreases. Thus the state
AC is more squeezed for a smaller value of $\del$. In the limit of
$\delta\to1$, the negativity $N_{\text{A:C}}$ vanishes since $a\to 1$,
implying that the purity of the state of mode $A$ approaches one.
\noindent
In terms of the symplectic eigenvalue $a$, the entanglement entropy of the mode A is given by
\begin{equation}
 S_\text{A}=\left(\frac{a+1}{2}\right)\log_2\left(\frac{a+1}{2}\right)-\left(\frac{a-1}{2}\right)\log_2\left(\frac{a-1}{2}\right).
\end{equation}
As the symplectic eigenvalue $a$ monotonically increases with the
scale factor, the entanglement entropy of the mode A also increases
with the scale factor. Thus the information shared between two modes A
and C increases because the mixedness of the state A grows. This
explains the delocalization of the information stored in A and the
spread of the partner's window function as visualized in
Fig. \ref{fig:partner-profile}. In the limit of $\del\rightarrow 1$,
$S_\text{A}$ approaches zero and A and C share no information.


\subsection{Purification of the bipartite Gaussian mode}\label{sec:partner_two_modes}
In this subsection, we construct partner modes CD for the two-mode
system
$\hat{\bs{\xi}}_\text{AB}=(\hat q_\text{A},\hat p_\text{A},\hat
q_\text{B},\hat p_\text{B})^T$.  See Fig. \ref{fig:purification} (b)
for the schematic picture of the setup.
To the authors' knowledge, an explicit formula to obtain the partner
modes of a two-mode system has not yet appeared in the
literature. Therefore, we explain here the derivation, although it is
quite similar to the arguments in the previous subsection, i.e., the
derivation of the partner formula for a one-mode system.

From the arguments in Sec.~\ref{sec:QIC}, the eight operators
$\hat{\bs{\xi}}_\text{AB}$ and $f_\psi(\hat{\bs{\xi}}_\text{AB})$, i.e.,
\begin{equation}
  \hat q_\text{A},\hat p_\text{A}, \hat q_\text{B}, \hat p_\text{B}, f_\psi(\hat q_\text{A}),
  f_\psi(\hat p_\text{A}), f_\psi(\hat q_\text{B}), f_\psi(\hat p_\text{B})
\end{equation}
define a system composed of four modes ABCD, which is in a pure state. We aim to construct two modes CD orthonormal to modes AB. 
Commutators between these operators are given by
\begin{align}
  &[\hat{\bs{\xi}}_\text{AB},\hat{\bs{\xi}}_\text{AB}^T]=i\bs{\Omega}_2,\quad [\hat{\bs{\xi}}_\text{AB},
    f_\psi(\hat{\bs{\xi}}_\text{AB}^T)]=i\bs{m}_\text{AB},\quad
    [f_\psi(\hat{\bs{\xi}}_\text{AB}),\hat{\bs{\xi}}_\text{AB}^T]=-i\bs{m}_\text{AB},\quad
    [f_\psi(\hat{\bs{\xi}}_\text{AB}),f_\psi(\hat{\bs{\xi}}_\text{AB}^T)]=i\bs{\Omega}_2,
\end{align}
where $\bs{m}_\text{AB}$ denotes the covariance matrix for the two
mode system $\hat{\bs{\xi}}_\text{AB}$:
\begin{align}
    \bs{m}_\text{AB}:=\expval{\{\hat{\bs{\xi}}_\text{AB},\hat{\bs{\xi}}^T_\text{AB}\}}=
    \expval{\{f_\psi(\hat{\bs{\xi}}_\text{AB}),f_\psi(\hat{\bs{\xi}}_\text{AB}^T)\}},\quad
  \expval{\left\{\hat{\bs{\xi}}_\text{AB},f_\psi(\hat{\bs{\xi}}_\text{AB}^T)\right\}}=-\bs{\Omega}_2.
\end{align}
To find modes orthogonal to the original mode
$\hat{\bs{\xi}}_\text{AB}$, we introduce operators
\begin{equation}
  \hat{\bs{\zeta}}:=f_\psi(\hat{\bs{\xi}}_\text{AB})-\bs{m}_\text{AB}\,
  \bs{\Omega}_2\,\hat{\bs{\xi}}_\text{AB}.
\end{equation}
They satisfy $[\hat{\bs{\zeta}},\hat{\bs{\xi}}_\text{AB}^T]=0$ since
\begin{align}
  [\hat{\bs{\zeta}},\hat{\bs{\xi}}_\text{AB}^T]
  &=[f_\psi(\hat{\bs{\xi}}_\text{AB}),\hat{\bs{\xi}}_\text{AB}^T]-[\bs{m}_\text{AB}\,\bs{\Omega}_2\,\hat{\bs{\xi}}_\text{AB},\hat{\bs{\xi}}_\text{AB}]
                           \notag\\
  &=-i\bs{m}_\text{AB}-\bs{m}_\text{AB}\,
    \bs{\Omega}_2[\hat{\bs{\xi}}_\text{AB},\hat{\bs{\xi}}_\text{AB}^T]\notag \\
  &=0
\end{align}
and therefore define modes orthogonal to the original modes AB. 
The commutators and the covariances for $\hat{\bs{\zeta}}$ are calculated as
\begin{equation}
  [\hat{\bs{\zeta}},\hat{\bs{\zeta}}^T]=i(\bs{\Omega}_2-\bs{m}_\text{AB}\,\bs{\Omega}_2\,\bs{m}_\text{AB}),\quad
  \expval{\left\{\hat{\bs{\zeta}},\hat{\bs{\zeta}}^T\right\}}=-\left(
    \bs{m}_\text{AB}+\bs{m}_\text{AB}\,\bs{\Omega}_2\,\bs{m}_\text{AB}\,\bs{\Omega}_2\,\bs{m}_\text{AB}\right).
\end{equation}
If $\bs{\Omega}_2=\bs{m}_\text{AB}\,\bs{\Omega}_2\,\bs{m}_\text{AB}$,
the two-mode system AB is in a pure state, implying their partners do
not exist. We therefore assume that
$\bs{\Omega}_2\neq \bs{m}_\text{AB}\,\bs{\Omega}_2\,\bs{m}_\text{AB}$.
We normalize $\hat{\bs{\zeta}}$ as
$\hat{\bs{\zeta}}=\bs{A}\,\hat{\bs{\xi}}_\text{CD}$ using a matrix
$\bs{A}$ so that the standard canonical commutation relation for modes
C and D
\begin{equation}
\quad[\hat{\bs{\xi}}_\text{CD},\hat{\bs{\xi}}_\text{CD}^T]
  =i\bs{\Omega}_2
\end{equation}
is satisfied. This is equivalent to a constraint on the
$4\times4$ matrix $\bs{A}$ given by
\begin{equation}
  \bs{A\Omega}_2\bs{A}^T=\bs{\Omega}_2-\bs{m}_\text{AB}\,\bs{\Omega}_2\,\bs{m}_\text{AB}.
\label{eq:relA}
\end{equation}
By using matrix $\bs{A}$ satisfying this condition, covariances of
normalized operators $\hat{\bs{\xi}}_\text{CD}$ are expressed as
\begin{align}
  \expval{\left\{\hat{\bs{\xi}}_\text{CD},\hat{\bs{\xi}}_\text{CD}^T\right\}}
  &=   \bs{A}^{-1}\expval{\left\{\hat{\bs{\zeta}},\hat{\bs{\zeta}}^T\right\}}(\bs{A}^{-1})^T
                                                             \notag \\
  &=-\bs{A}^{-1}(\bs{m}_\text{AB}+\bs{m}_\text{AB}\,\bs{\Omega}_2\,\bs{m}_\text{AB}\,\bs{\Omega}_2\,\bs{m}_\text{AB})
    (\bs{A}^{-1})^T=:\bs{m}_{\text{CD}}, \label{eq:cov1}\\
  \expval{\left\{\hat{\bs{\xi}}_\text{CD},\hat{\bs{\xi}}_\text{AB}^T\right\}}
  &=
    \bs{A}^{-1}(\bs{\Omega}_2-\bs{m}_\text{AB}\,\bs{\Omega}_2\,\bs{m}_\text{AB})=:\bs{m}_{\text{AB,CD}}^T
    , \label{eq:cov2}
  \\
  \expval{\left\{\hat{\bs{\xi}}_\text{AB},\hat{\bs{\xi}}_\text{CD}^T\right\}}
  &= \left[\bs{A}^{-1}(\bs{\Omega}_2-\bs{m}_\text{AB}\,\bs{\Omega}_2\,\bs{m}_\text{AB})\right]^T
    =:\bs{m}_{\text{AB,CD}}, \label{eq:cov3}
  \\
  \expval{\left\{\hat{\bs{\xi}}_\text{AB},\hat{\bs{\xi}}_\text{AB}^T\right\}}
  &=:\bs{m}_{\text{AB}}.
    \label{eq:cov4} 
\end{align}
These covariances define a state for the four-mode system ABCD as
\begin{equation}
  \bs{m}_\text{ABCD}=
  \begin{bmatrix}
    \bs{m}_\text{AB}&\bs{m}_\text{AB,CD}\\\bs{m}_\text{AB,CD}^T&\bs{m}_\text{CD}
  \end{bmatrix}\label{eq:cov_mat_4}.
\end{equation}
Since the four-mode system ABCD is in a pure state, the purity
condition is satisfied
\begin{equation}
\bs{m}_\text{ABCD}\,\bs{\Omega}_4
  \,\bs{m}_\text{ABCD}=
  \bs{\Omega}_4,
\end{equation}
where $\bs{\Omega}_4=\bigoplus_{i=1}^4\bs{J}$. In this case, the state of the total system is decomposed as
\begin{equation}
  \ket{\psi}=\ket{\psi'}_\text{ABCD}\otimes\ket{\psi''}_{\overline{\text{ABCD}}},
\end{equation}
where $\ket{\psi'}_\text{ABCD}$ denotes a pure state of the four-mode
system ABCD defined by the covariance matrix $\bs{m}_{\text{ABCD}}$,
while $\ket{\psi}_{\overline{\text{ABCD}}}$ is a pure state for its
complement system $\overline{\text{ABCD}}$. This decomposition implies
that there is no correlation between the four-mode system ABCD and its
complement $\overline{\text{ABCD}}$. Therefore, all the information on
the correlation between AB and its complement is confined to the
four-mode system ABCD.

We look for the matrix $\bs{A}$ satisfying Eq.~\eqref{eq:relA} by
using the standard form of the covariance matrix of a two-mode
Gaussian state \cite{duan2000inseparability,Serafini2007}. Our aim is
to find the partner modes of local modes A and B defined at spatial
points $x_\text{A}$ and $x_\text{B}$. Because of the spatial translation
symmetry, the symplectic eigenvalue of the covariance matrix of mode A
is equal to that of B. Therefore, without loss of generality, we can
assume that the covariance matrix of the bipartite system AB is given
by the standard form of symmetric Gaussian state
\begin{equation}
  \bs{m}_\text{AB}=
  \begin{bmatrix}
    a&0&d_1&0\\ 0&a&0&d_2\\ d_1&0&a&0\\0&d_2&0&a
  \end{bmatrix},
  \label{eq:cov-standard}
\end{equation}
after performing a local symplectic transformation on each mode.

Using the standard form of $\bs{m}_\text{AB}$,
the right-hand side of Eq. \eqref{eq:relA} is expressed as
\begin{equation}
  \bs{\Omega}_2-\bs{m}_\text{AB}\,\bs{\Omega}_2\,\bs{m}_\text{AB}=
  \begin{bmatrix}
    (1-a^2-d_1d_2)\bs{J}&-a(d_1+d_2)\bs{J}\\
    * & (1-a^2-d_1d_2)\bs{J}
  \end{bmatrix}
.
\end{equation}
As noted in the previous section, the solution $\bs{A}$ of
Eq. \eqref{eq:relA} is not uniquely determined as there is
remaining freedom for fixing canonical operators for CD. As an ansatz
for the matrix $\bs{A}$, we adopt
\begin{equation}
  \bs{A}=
  \begin{bmatrix}
    g \bs{X}&h \bs{X}\\ h \bs{X}&g \bs{X}
  \end{bmatrix}.
\end{equation}
Because
\begin{equation}
    \bs{A}\bs{\Omega}_2\bs{A}=
    -
    \begin{bmatrix}
      (g^2+h^2)\bs{J}& 2gh \bs{J}\\
      2gh \bs{J}&(g^2+h^2)\bs{J}
    \end{bmatrix},
\end{equation}
the constraint in Eq.~\eqref{eq:relA} is equivalent to
\begin{equation}
  g^2+h^2=a^2+d_1d_2-1,\quad 2gh=a(d_1+d_2),
\end{equation}
and the solution is
\begin{equation}
   g=\frac{1}{2}(\sqrt{x+y}+\sqrt{x-y}),\quad h=\frac{1}{2}(\sqrt{x+y}-\sqrt{x-y}),
\end{equation}
where we introduced
\begin{equation}
  x=a^2+d_1d_2-1,\quad y=a(d_1+d_2).
  \label{eq:xy}
\end{equation}
Because the inverse of $\bs{A}$ is given by
\begin{equation}
  \bs{A}^{-1}=\frac{1}{g^2-h^2}
  \begin{bmatrix}
    g \bs{X}&-h \bs{X}\\ -h \bs{X}&g \bs{X}
  \end{bmatrix}, 
\end{equation}
the covariance matrix of the pure state of the four-mode system ABCD
is obtained from \eqref{eq:cov1}-\eqref{eq:cov4} as
\begin{align}
  &\bs{m}_\text{ABCD}=
  \begin{bmatrix}
    \bs{m}_\text{AB}&\bs{m}_\text{AB,CD}\\\bs{m}_\text{AB,CD}^T&\bs{m}_\text{CD}
  \end{bmatrix}, \label{eq:4mode1}\\
    &\bs{m}_\text{AB}=
  \begin{bmatrix}
    a&0&d_1&0\\
    0&a&0&d_2\\
    d_1&0&a&0\\
    0&d_2&0&a
  \end{bmatrix},\quad
        \bs{m}_\text{CD}=
               \begin{bmatrix}
                 a&0&d_2&0\\0&a&0&d_1\\d_2&0&a&0\\0&d_1&0&a
               \end{bmatrix}
               ,\quad
   \bs{m}_\text{AB,CD}=
    \begin{bmatrix}
      g&0&h&0\\
      0&-g&0&-h\\
      h&0&g&0\\
      0&-h&0&-g
    \end{bmatrix}. \label{eq:4mode2}
\end{align}
The canonical operators describing partner modes CD are given by 
\begin{align}
  \hat{\bs{\xi}}_{\text{CD}}
  &=\bs{A}^{-1}(f_\psi(\hat{\bs{\xi}}_\text{AB})-\bs{m}_\text{AB}\,\bs{\Omega}_2\,\hat
    {\bs{\xi}}_\text{AB}) \notag \\
  &=
\frac{1}{g^2-h^2}
\begin{bmatrix}
  g\bs{X}& -h\bs{X}\\ -h\bs{X}&g\bs{X}
\end{bmatrix}
(f_\psi(\hat{\bs{\xi}}_\text{AB})-\bs{m}_\text{AB}\,\bs{\Omega}_2\,\hat
  {\bs{\xi}}_\text{AB}),\label{eq:partner_formula_two}
\end{align}
which establishes the partner formula for a two-mode symmetric
Gaussian state. Note that as $f_\psi(\hat{\bs{\xi}}_\text{B})$ is
obtained by replacing $x_\text{A}\rightarrow x_\text{B}$ in
$f_\psi(\hat{\bs{\xi}}_\text{A})$, their window functions are
calculated by shifting the window functions of
$f_\psi(\hat{\bs{\xi}}_\text{A})$ given in Eqs.~\eqref{eq:partner-f1}
and \eqref{eq:partner-f2}. Therefore, their behaviors are expressed by
Fig.~\ref{fig:partner-profile}, except for the shift of the
centers. Since Eq.\eqref{eq:partner_formula_two} implies that the
window functions of the partner CD of AB are expressed by operators
given by $\hat{\bs{\xi}}_\text{AB}$ and
$f_\psi(\hat{\bs{\xi}}_\text{AB})$, we find that they are given by
linear combinations of functions which are localized around
$x_\text{A}$ and $x_\text{B}$, whose tails change depending on the
scale factors $a_{\text{sc}}$.

  %

\section{Monogamy and Separability}\label{sec:4}
We regard the bipartite system AB as a subsystem embedded in the pure
four-mode state ABCD. Then an entanglement measure
$\widetilde E(\text{A:B})$ between A and B (internal entanglement) and
an entanglement measure $E(\text{AB:CD})$ between AB and CD (external
entanglement) are expected to obey the following monogamy inequality
\cite{Camalet2017,Camalet2017a,Camalet2018,Zhu2020,Zhu2021}:
\begin{equation}
  \widetilde E(\text{A:B})+E(\text{AB:CD})\le\widetilde E_\text{max},
  \label{eq:mono-camalet}
\end{equation}
where $\widetilde E_\text{max}$ is the maximum of
$\widetilde E(\text{A:B})$. This inequality represents a trade-off
relation between internal and external entanglement and has been
proven to hold for finite-dimensional Hilbert space cases, including qubit
systems.
For qubit cases, explicit
forms of inequalities are presented in terms of various entanglement measures
(concurrence, entanglement of formation, and negativity).
In this paper, based on the specific representation of the four-mode Gaussian state \eqref{eq:4mode1} and \eqref{eq:4mode2} which
purifies AB, we show this type of monogamy inequality also holds for
Gaussian states.

\subsection{Parametrization of the bipartite Gaussian state}
In the standard form, the covariance matrix \eqref{eq:cov-standard} of
the bipartite symmetric Gaussian state AB includes three parameters
$a,d_1,d_2$. For later convenience, we parametrize it with $a,x,y$
where $x,y$ are defined by \eqref{eq:xy}. By solving Eq. \eqref{eq:xy}
with respect to $d_1,d_2$, we get
\begin{equation}
  d_1=\frac{y}{2a}+\sqrt{\frac{y^2}{4a^2}-(x-a^2+1)},\quad
  d_2=\frac{y}{2a}-\sqrt{\frac{y^2}{4a^2}-(x-a^2+1)},
\end{equation}
and $d_1-d_2=\sqrt{y^2/a^2-4(x-a^2+1)}\ge 0,\quad d_1d_2=x-a^2+1$.
Thus $d_1$ and $d_2$ are expressed using $a,x,y$.  Symplectic
eigenvalues of the covariance matrix $\bs{m}_\text{AB}$ are given by
\begin{align}
  &\nu_1^2=(a+d_1)(a+d_2)=x+y+1\ge
    1,\quad\nu_2^2=(a-d_1)(a-d_2)=x-y+1\ge 1,
\end{align}
and
$\mathrm{det}\,\bs{m}_\text{AB}=(a^2-d_1^2)(a^2-d_2^2)=\nu_1^2\nu_2^2=\tilde\nu_1^2\tilde\nu_2^2=(x+1)^2-y^2$.
Symplectic eigenvalues of the partially transposed covariance matrix
$\widetilde{\bs{m}}_\text{AB}$ are expressed as
\begin{align}
  &\tilde\nu_{1}^2=(a+d_1)(a-d_2)=2a^2-x-1+\sqrt{y^2-4a^2(x-a^2+1)},\\
  &\tilde\nu_2^2=(a-d_1)(a+d_2)=2a^2-x-1-\sqrt{y^2-4a^2(x-a^2+1)}.
\end{align}
The sum of these symplectic eigenvalues satisfies $\nu_1^2+\nu_2^2+\tilde\nu_1^2+\tilde\nu_2^2=4a^2$.
For real values of $d_{1}$ and $d_2$, it holds
\begin{equation}
  y^2\ge4a^2(x-a^2+1).
\end{equation}
The modes A and B are entangled if $0<\tilde\nu_2<1<\tilde\nu_1$, or equivalently, $(x+2)^2-y^2<4a^2$. The negativity of the state AB is given by
\begin{equation}
  N_\text{A:B}=\frac{1}{2}\mathrm{max}\left[\frac{1}{\tilde\nu_2}-1,0\right].
  \label{eq:negAB}
\end{equation}

\begin{figure}[H] 
  \centering
  \includegraphics[width=0.4\linewidth]{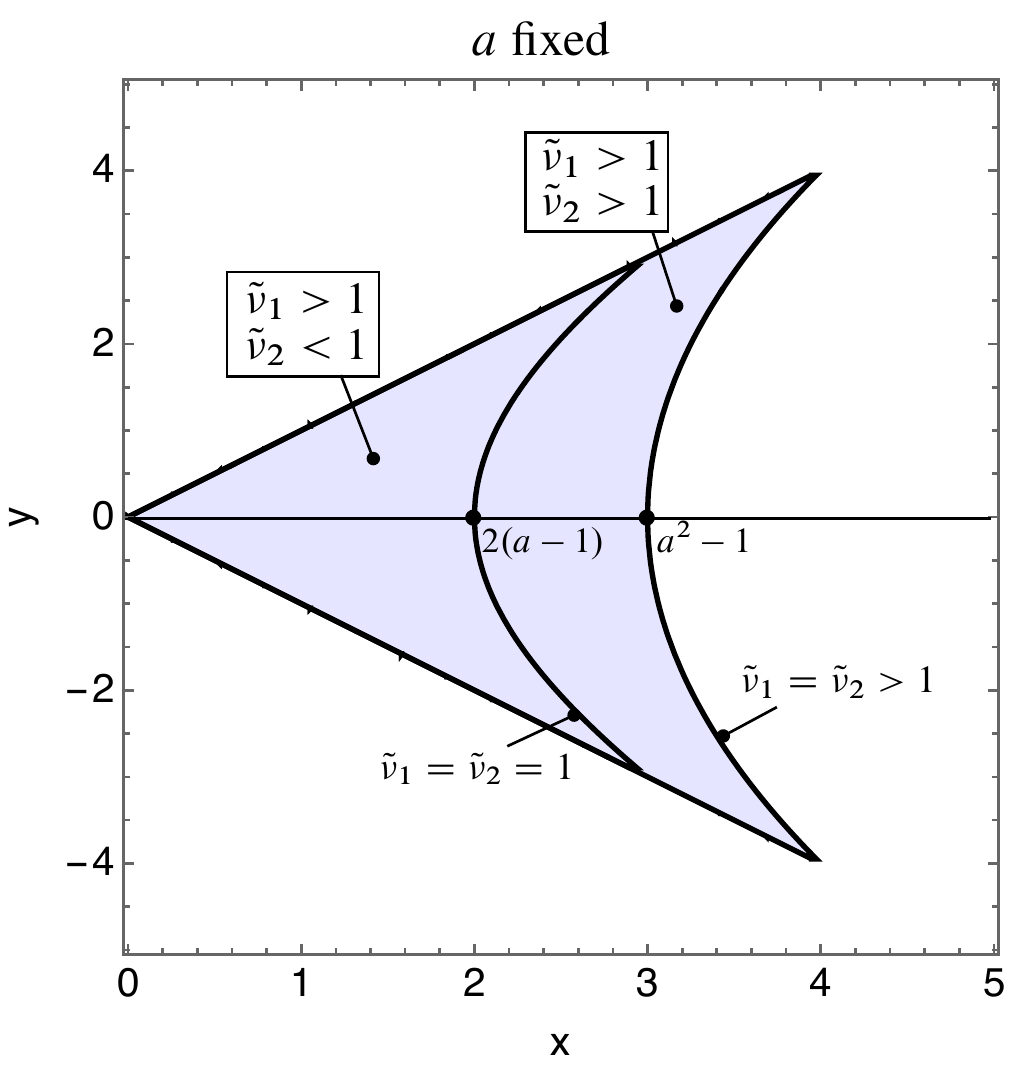}
  \caption{The parameter region representing bipartite Gaussian states
    in the $(x,y)$ plane (shaded region). The region
    $(x+2)^2-y^2<4a^2$ corresponds to entangled states and the region
    $(x+2)^2-y^2>4a^2$ corresponds to separable states.}
  \label{fig:xy-region}
\end{figure}

For a fixed $a$, the minimum of $\tilde\nu_2$ is attained at $x=y=0$ and given by
\begin{equation}
  \tilde\nu_2|_\text{min}=a-\sqrt{a^2-1}.
\end{equation}
In this case,  the bipartite state AB is a two-mode squeezed pure
state with a squeezing parameter $r=\cosh^{-1}a$, and its covariance
matrix is given by Eq. \eqref{eq:cov-standard} with
$d_1=-d_2=\sqrt{a^2-1}$.  Thus the maximum of $N_\text{A:B}$ is
\begin{equation}
  N_\text{A:B}|_\text{max}=\frac{1}{2}\left(a-1+\sqrt{a^2-1}\right). \label{eq:Nmax}
\end{equation}
The bipartite state AB becomes separable at $\tilde\nu_2=1$, and this condition is equivalent to
\begin{equation}
  (x+2)^2-y^2=4a^2.
\end{equation}
With a fixed value of $a$, it is possible to draw a parameter region
in the $(x,y)$ plane where $\bs{m}_\text{AB}$ represents a physical
Gaussian state (Fig. \ref{fig:xy-region}). The region is bounded by
$x=|y|$ corresponding to $\nu=1$ (positivity of the state) and
$y^2=4a^2(x-a^2+1)$ corresponding to the reality condition of
$d_{1,2}$. The region is divided into two regions: one corresponds to
entangled state, and the other corresponds to separable states. A pure
state is located at $x=y=0$, corresponding to a two-mode squeezed pure
state. The state becomes separable for $a=1$.

Symplectic eigenvalues of the four-mode state ABCD are given by
\begin{align}
  &\sqrt{(a-d_1)(a-d_2)-(g-h)^2}=1,\quad \sqrt{(a+d_1)(a+d_2)-(g+h)^2}=1,
\end{align}
which implies the state ABCD is pure.  Symplectic eigenvalues of the
partially transposed state with bipartition AB:CD are given by
\begin{align}
  (\tilde\nu_{2\pm})^2&=(a-d_1)(a-d_2)+(g-h)^2\pm2|g-h|\sqrt{(a-d_1)(a-d_2)}
                        \notag \\
  &=(\sqrt{x-y+1}\pm\sqrt{x-y})^2 , \\
  (\tilde\nu_{1\pm})^2&=(a+d_1)(a+d_2)+(g+h)^2\pm2|g+h|\sqrt{(a+d_1)(a+d_2)}
                        \notag \\
  &=(\sqrt{x+y+1}\pm\sqrt{x+y})^2.
\end{align}
Therefore, $ \tilde\nu_{2\pm}=\nu_2\pm \sqrt{\nu_2^2-1},
  \tilde\nu_{1\pm}=\nu_1\pm
  \sqrt{\nu_1^2-1}$, 
and the negativity for the bipartition AB:CD is calculated as
\begin{align}
  N_\text{AB:CD}&=\frac{1}{2}\left(\nu_1+\sqrt{\nu_1^2-1}\right)
    \left(\nu_2+\sqrt{\nu_2^2-1}\right)-\frac{1}{2}>0.
\end{align}

\subsection{Monogamy relation for Gaussian states}
We examine the monogamy inequality \eqref{eq:mono-camalet} for
Gaussian states.  For the qubit case treated in \cite{Camalet2017}, as
the entanglement measure in this monogamy inequality,
$\widetilde E(\text{A:B})$ can be the negativity $N_\text{A:B}$.
$E(\text{AB:CD})$ is a decreasing function of the negativity
$N_\text{AB:CD}$, and the explicit form of this function is presented
in \cite{Camalet2017}.  In the present analysis with Gaussian states,
we also adopt negativity as an entanglement measure to show a monogamy
inequality.  

\begin{figure}[H]
  \centering
  \includegraphics[width=1\linewidth]{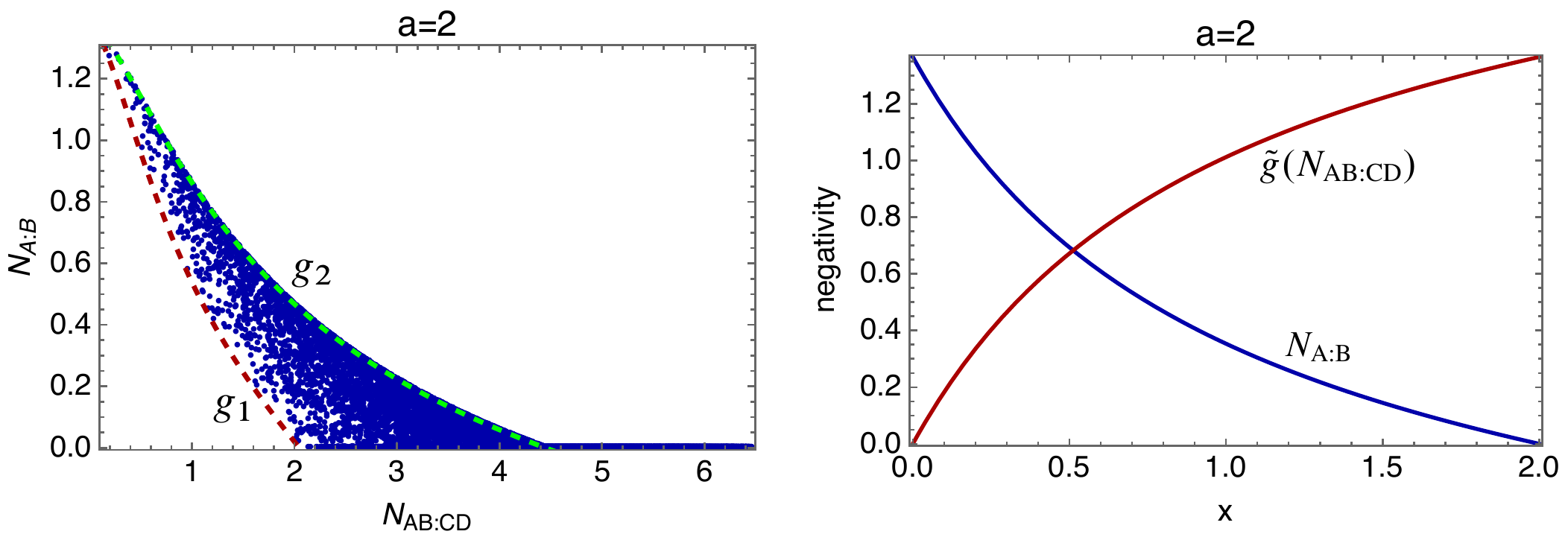}
  \caption{Left panel: distribution of  $(N_\text{AB:CD},N_\text{A:B})$ for 
    randomly generated bipartite Gaussian states with fixed $a$ [11189 sets of parameters $(x,y)$]. States are
    located in the region surrounded by the  dashed green line
    (corresponds to $y=0$) and the dashed red line (corresponds to
    $|y|=x$). In this case with $a=2$, for states with
    $\beta=2+\sqrt{6}< N_\text{AB:CD}$, $N_\text{A:B}=0$; thus the
    external correlation between AB and CD limits the amount of the internal
    entanglement between A and B. The same relation also holds for any
    values of $a>1$. Right panel: Behavior of $N_\text{A:B}$ and $\tilde g(N_\text{AB:CD})$
    as functions of $x$ (with $y=0$). A trade-off relation between the
    internal entanglement $N_\text{A:B}$ and the external entanglement
  $N_\text{AB:CD}$ can be observed. For $\beta=2(a-1)\le x$, $N_\text{A:B}=0$.}
  \label{fig:random}
\end{figure}

As we have already presented, negativities $N_\text{A:B}$ and
$N_\text{AB:CD}$ are expressed as functions of $a,x,y$:
\begin{align}
 &N_\text{A:B}(x,y,a)=\frac{1}{2}\left(\frac{1}{\tilde\nu_2}-1\right),\quad
  \tilde\nu_{2}^2=2a^2-x-1-\sqrt{y^2-4a^2(x-a^2+1)}, \\
  &N_\text{AB:CD}(x,y)=\frac{1}{2}\left(\sqrt{x+y+1}+\sqrt{x+y}\right)
  \left(\sqrt{x-y+1}+\sqrt{x-y}\right)-\frac{1}{2}.
\end{align}
To capture qualitative behavior of the monogamy relation between
$N_\text{A:B}$ and $N_\text{AB:CD}$, we randomly generate parameters
$x,y$ of bipartite Gaussian states with fixed $a$.  The left panel of
Fig. \ref{fig:random} shows the distribution of
$(N_\text{AB:CD},N_\text{A:B})$ for randomly generated bipartite
Gaussian states. We observe that all bipartite Gaussian states are
confined in a region surrounded by lines
$N_\text{A:B}=g_1(N_\text{AB:CD}), N_\text{A:B}=g_2(N_\text{AB:CD})$,
and $N_\text{A:B}=0$, i.e.,
\begin{equation}
  \begin{cases}
    g_1(N_\text{AB:CD})\le N_\text{A:B}\le g_2(N_\text{AB:CD})\quad
    &         \text{for}\quad0\le N_\text{AB:CD}\le \al, \\
    0\le N_\text{A:B}\le g_2(N_\text{AB:CD})\quad
    & \text{for}\quad \al\le N_\text{AB:CD}\le \beta, \\
     N_\text{A:B}=0\quad &\text{for}\quad \beta\le N_\text{AB:CD},
  \end{cases}
  \label{eq:mono-gauss1}
\end{equation}
where functions $g_1$ and $g_2$ define the relations between
$N_\text{A:B}$ and $N_\text{AB:CD}$ on $|y|=x$ and $y=0$,
respectively. They are monotonically decreasing functions of
$N_\text{AB:CD}$. The parameters $\al$ and $\beta$ are defined by
$g_1(\al)=0$ and $g_2(\beta)=0$.  When $N_\text{A:B}$ attains its
maximum for a fixed $a$, $N_\text{AB:CD}=0$, and hence, the
bipartite state AB is pure. The explicit expression of functions $g_2$
and $\beta$ are obtained as
\begin{align}
  &g_2=\frac{1}{2}\left(-1+\left(a-\sqrt{a^2-\frac{(N_\text{AB:CD}+1)^2}{2N_\text{AB:CD}+1}}\right)^{-1}\right),\\
  &\beta=-\frac{1}{2}+\left(\sqrt{a-1/2}+\sqrt{a-1}\right)^2.
\end{align}
\noindent
For $\al\le N_\text{AB:CD}\le\beta$ with fixed $a$, the following  inequality holds:
\begin{equation}
  N_\text{A:B}\le g_2(N_\text{AB:CD},a).
  \label{eq:mono-gauss1b}
\end{equation}
Thus the function $g_2$ determines the upper bound of $N_\text{A:B}$
for given values of $N_\text{AB:CD}$ and $a$. Here, $g_2$ is a decreasing function of $N_\text{AB:CD}$ and becomes zero at $N_\text{AB:CD}=\beta$. 
We rewrite this
inequality as
\begin{equation}
  N_\text{A:B}+\tilde g(N_\text{AB:CD},a)\le N_\text{A:B}|_\text{max}(a),
  \label{eq:mono-gauss2}
\end{equation}
where we introduced
\begin{equation}
    \tilde{g}(N_\text{AB:CD},a):=
    \begin{cases}
        0\quad& (N_\text{AB:CD}=0)\\
        N_\text{A:B}|_\text{max}(a)-g_2(N_\text{AB:CD},a) \quad &(0\leq N_\text{AB:CD}\leq \beta)\\
        N_\text{A:B}|_\text{max}(a)\quad &( \beta\leq N_\text{AB:CD})
    \end{cases}
\end{equation}
and $N_\text{A:B}|_\text{max}$ is defined by \eqref{eq:Nmax}. Note
that for fixed $a$, $\tilde g$ is a non-negative monotonically
increasing function of $N_\text{AB:CD}$. Furthermore, it vanishes if
$N_\text{AB:CD}=0$. In this sense, the function $\tilde g$ defines an
entanglement measure for bipartition AB:CD for each $a$. Thus for
Gaussian states with fixed $a$, we have obtained the monogamy
inequality \eqref{eq:mono-gauss2} that represents a trade-off relation
between the internal entanglement $N_\text{A:B}$ and the external
entanglement $\tilde{g}(N_\text{AB:CD},a)$. When
$\beta\leq N_\text{AB:CD}$, $\tilde{g}(N_\text{AB:CD},a)$ attains its
maximum, and the negativity between A and B automatically vanishes,
i.e., $N_\text{A:B}=0$.


The right panel of Fig. \ref{fig:random} shows the behavior of
$N_\text{A:B}$ and $\tilde g(N_\text{AB:CD})$ as functions of $x$ when
$y=0$. This case corresponds to saturation of the inequality
\eqref{eq:mono-gauss2} and the following equality holds:
\begin{equation}
  N_\text{A:B}+\tilde g(N_\text{AB:CD},a)= N_\text{A:B}|_\text{max}(a).
  \label{eq:mono-eq}
\end{equation}


\subsection{Monogamy for local modes in the de Sitter universe}
\begin{figure}[H]
  \centering
  \includegraphics[width=1\linewidth]{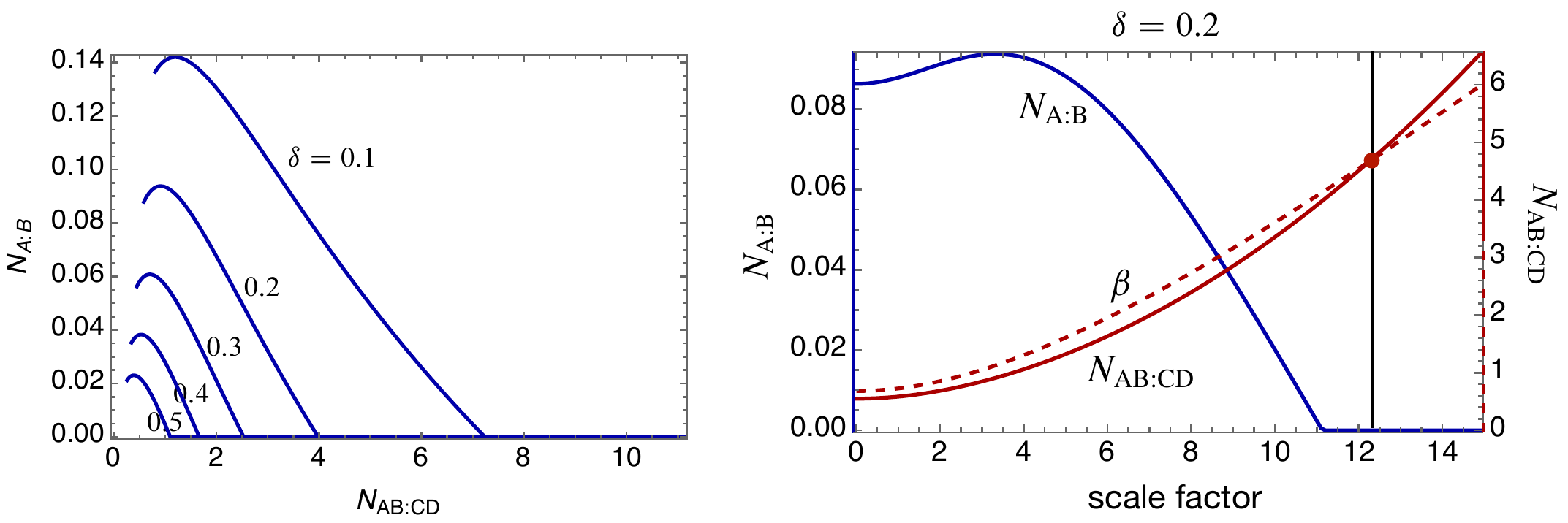}
  \caption{Left panel: relation between $N_\text{A:B}$ and
    $N_\text{AB:CD}$ for fixed values of $\delta$. Right panel:
    evolution of $N_\text{A:B}$, $N_\text{AB:CD}$ and $\beta$ as
    functions of the scale factor. The solid red circle denotes the
    location at $N_\text{AB:CD}=\beta$ and the solid black line
    indicates a value of the scale factor. For the right side of this
    point, $\beta<N_\text{AB:CD}$ and the monogamy inequality
    \eqref{eq:mono-gauss1c} implies $N_\text{A:B}=0$.}
  \label{fig:n2n4-DS}
\end{figure}

For the scalar field in the de Sitter universe, components of the
covariance matrix of local Gaussian modes are functions of
$a_\text{sc}$ and $\del$.  Figure \ref{fig:n2n4-DS} shows the
evolution of $N_\text{A:B}$ and $N_\text{AB:CD}$ with fixed
$\del$. The left panel shows relations between $N_\text{A:B}$ and
$N_\text{AB:CD}$ with different values of $\del$. The state evolves
from $a_\text{sc}=0$ that corresponds to the left edges of each
line. As we have already observed, $N_\text{A:B}$ becomes zero when
the physical size $\del\times a_\text{sc}$ of local modes exceeds the
Hubble horizon scale $H^{-1}$. On the other hand, $N_\text{AB:CD}$
increases monotonically with the scale factor for a fixed value of
$\del$; thus, $N_\text{A:B}$ becomes zero as $N_\text{AB:CD}$ reaches
a some critical value.

The right panel of Fig. \ref{fig:n2n4-DS} shows the evolution of
negativities and $\beta$ as functions of the scale factor for
$\delta=0.2$. The behavior of $N_\text{A:B}$ and $N_\text{AB:CD}$
represents a trade-off relation between them. From the argument in the
previous subsection, they satisfy the monogamy relation
\begin{equation}
  N_\text{A:B}+\tilde g(N_\text{AB:CD},a(a_\text{sc},\del))\le N_\text{A:B}|_\text{max}(a(a_\text{sc},\del)).
  \label{eq:mono-gauss1c}
\end{equation}
Note that this inequality is essentially the same as
\eqref{eq:mono-gauss2}, but the parameter $a$ becomes a function of
$a_\text{sc}$ and $\del$.  It explains the separable behavior of the
bipartite system AB as a monogamy relation between internal
entanglement and external entanglement; for
$\beta(a(a_\text{sc},\del))\leq N_\text{AB:CD}$, the function
$\tilde g$ attains its maxima while $N_\text{A:B}$ vanishes. Thus,
this inequality provides a sufficient condition of separability for
the bipartite system AB. Although $N_\text{A:B}$ becomes zero before
$N_\text{AB:CD}$ reaches $\beta$ (see right panel of
Fig. \ref{fig:n2n4-DS}), this behavior is consistent with
\eqref{eq:mono-gauss1c}.  The tightness of the monogamy inequality
depends on the parameter $\del$ in the present setup. Actually, as
$\del$ increases, the difference between
$g_2(N_\text{AB:CD},a(a_\text{sc},\del))$ and $N_\text{A:B}$
decreases. In the limit of $\del\rightarrow 1$ (pure state limit),
$N_\text{A:B}=g_2(N_\text{AB:CD},a)$ holds because
$N_\text{A:B}\rightarrow 0$ and $N_\text{AB:CD}\rightarrow 0$, which
implies that equality in \eqref{eq:mono-gauss1c} trivially holds.

\section{Summary and conclusion}\label{sec:5}

We investigated the emergence of separability for local bipartite
modes assigned to two spatial regions in the de Sitter universe. The
bipartite mode AB becomes separable after their separation exceeds the
Hubble horizon scale. To understand the emergence of this separability
from the viewpoint of entanglement monogamy, we considered
purification of the local mode AB and obtained the pure four-mode
state ABCD applying the partner formula. Then, we found the monogamy
inequality between the negativity $N(\text{A:B})$ and
$N(\text{AB:CD})$ for the four-mode Gaussian state, which is an
extension of Camalet's monogamy relation to continuous variable
systems. It is demonstrated that the separability of the mode AB can
be understood as the monogamy property between the internal and the
external entanglements, and the monogamy inequality provides a
sufficient condition for the separability of the local mode AB defined
from the quantum field. In the stochastic approach to inflation
\cite{Starobinsky1986}, local oscillator modes are defined as long
wavelength components of the inflaton field. The introduced local
modes are treated as ``classical" stochastic variables, and they obey
a Langevin equation with a stochastic noise originating from the short
wavelength quantum fluctuations. Although the stochastic approach to
inflation is a phenomenological treatment of quantum fields in the de
Sitter spacetime and is widely employed to investigate the physics
related to cosmic inflation, its justification is still missing. Our
investigation of this paper provides one reasoning to this method from
the viewpoint of quantum information; local modes in the de Sitter
universe lose quantum correlation when their separation exceeds the
cosmological horizon, and this behavior is related to delocalization
of partner modes.

The partner formula adopted in this study may provide a new
perspective on information sharing in multipartite quantum
systems. Indeed, as shown in Fig. \ref{fig:partner-profile}, the
spatial profiles of partner modes can be visualized and they are
helpful in capturing how the information of a system is shared with
its partners. The information stored in a system is lost but classical
properties of the system appear as a result of decoherence via
information sharing with its partners (environments). This direction
of investigation is closely related to the concept of ``quantum
Darwinism" \cite{Zurek2009a} which states that the
emergence of a classical
behavior of a quantum system, such as objectivity, is connected with
the amount of information of the system redundantly shared or stored
in the environment. Thus spatial profiles of partner modes of the
system may  help to quantify this redundancy of the information
and to understand the quantum to classical transition in the early
universe.

\begin{acknowledgements}
  We  thank A. Matsumura for providing his valuable
  insight on the subject. This research was supported in part by a
  Grant-in-Aid for Scientific Research No. 19K03866, No. 22H05257 and
  No. 23H01175 (Y.N.) from the Ministry of Education, Culture, Sports,
  Science, and Technology (MEXT), Japan. K.Y. acknowledges support
  from the JSPS Overseas Research Fellowships.
\end{acknowledgements}

\appendix
\section{\label{sec:appendix}CONVENTIONAL MONOGAMY RELATION}
\begin{figure}[H]
  \centering
  \includegraphics[width=1\linewidth]{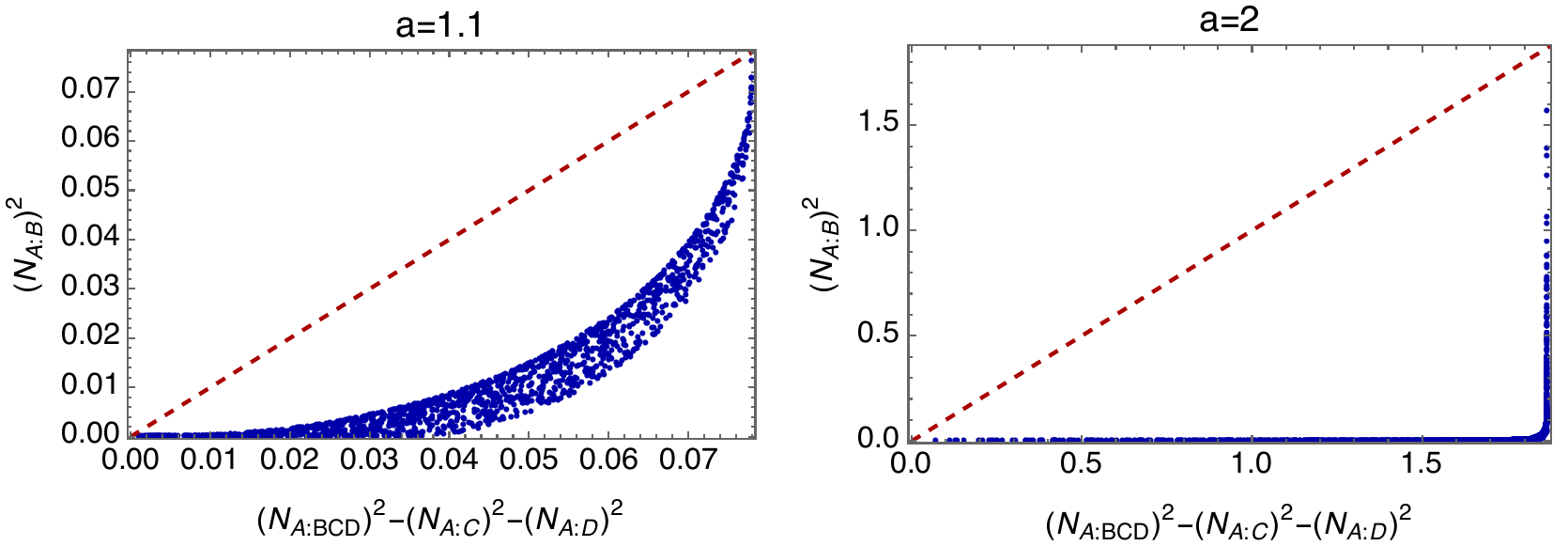}
  \caption{Demonstration of the monogamy inequality
    \eqref{eq:mon-conv} for the four-mode Gaussian state with the
    covariance matrix \eqref{eq:4mode1}. The number of randomly
    generated states is 1500. As an entanglement measure, the square
    of negativity is adopted. All generated states are located below
    the dashed red line, which represents the equality of the relation
    \eqref{eq:mon-conv}.}
  \label{fig:mono-conv}
\end{figure}

We present the conventional monogamy relation for Gaussian states \cite{Hiroshima2007,Adesso2007a,Adesso2014}. For the four-mode pure Gaussian state  ABCD with the covariance matrix \eqref{eq:4mode1},
\begin{equation}
E(\text{A:B})+E(\text{A:C})+E(\text{A:D})\le E(\text{A:BCD}),
\label{eq:mon-conv}
\end{equation}
where $E$ denotes a suitably chosen entanglement measure and this inequality  holds with the square of negativity or square of logarithmic negativity as entanglement measures. We demonstrate it for randomly generated Gaussian states  by taking $E$ as the square of negativity. Negativities are given by
\begin{align}
&N_\text{A:BCD}=\frac{1}{2}\left(a+\sqrt{a^2-1}-1\right),\\
&N_\text{A:C}=\mathrm{max}\left[\frac{1}{2a-(\sqrt{x+y}+\sqrt{x-y})}-\frac{1}{2},0\right],\\
&N_\text{A:D}=\mathrm{max}\left[\frac{1}{2a-(\sqrt{x+y}-\sqrt{x-y})}-\frac{1}{2},0\right],
\end{align}
and $N_\text{A:B}$ is given by \eqref{eq:negAB}.  We can observe the
monogamy inequality \eqref{eq:mon-conv} indeed holds for this
four-mode Gaussian state (Fig. \ref{fig:mono-conv}) because generated
states are located below the  dashed red lines that represent equality
of \eqref{eq:mon-conv}. However, Fig. \ref{fig:mono-conv} shows that
states deviate from the dashed red lines as the parameter $a$
increases. Therefore, the monogamy inequality \eqref{eq:mon-conv} does
not provide a useful tight constraint on the separability of the
bipartite state
AB. 


%

\end{document}